\documentstyle[preprint,aps,eqsecnum]{revtex}
\input psfig.sty
\begin{document}
\draft 
\preprint{IFT-HET-96-20,OSU-TA-20/96}
\baselineskip.5cm
\parskip4pt
\date{September 3, 1996}

\title{The secondary infall model of galactic halo formation
and the spectrum of cold dark matter particles on Earth.}

\author{P. Sikivie$^{1}$, I. I.\ Tkachev$^{2,3}$ and Yun Wang$^{4}$}
\address{
$^{1}$Department of Physics, University of Florida, Gainesville, FL 32611}
\address{
$^{2}$Department of Physics, The Ohio State University,
Columbus, OH 43210}
\address{
$^{3}$Institute for Nuclear Research,
Russian Academy of Sciences, Moscow 117312, Russia}
\address{
$^{4}$NASA/Fermilab Astrophysics Center,
FNAL, Batavia, IL~~60510}

\maketitle

\begin{abstract}
\baselineskip0.5cm
The spectrum of cold dark matter particles on Earth is expected to have 
peaks in velocity space associated with particles which are falling onto
the Galaxy for the first time and with particles which have fallen in 
and out of the Galaxy only a small number of times in the past.  We 
obtain estimates for the velocity magnitudes and the local densities of 
the particles in these peaks. To this end we use the secondary infall 
model of galactic halo formation which we have generalized to take account 
of the angular momentum of the dark matter particles.  The new model is 
still spherically symmetric and it admits self-similar solutions.  
In the absence of angular momentum, the model produces flat rotation 
curves for a large range of values of a parameter $\epsilon$ which 
is related to the spectrum of primordial density perturbations.  We find 
that the presence of angular momentum produces an effective core radius, 
i.e. it makes the contribution of the halo to the rotation curve go to 
zero at zero radius.  The model provides a detailed description of the 
large scale properties of galactic halos including their density profiles, 
their extent and total mass.  We obtain predictions for the kinetic 
energies of the particles in the velocity peaks and estimates for their 
local densities as functions of the amount of angular momentum, the age 
of the universe and $\epsilon$.  
\end{abstract}

\narrowtext

\section{Introduction}
\label{sec:in}

Experiments are presently under way which attempt to identify the nature 
of dark matter \cite{dmr} by direct detection on Earth. The dark matter 
candidates which are being searched for in this manner are WIMPs and 
axions. WIMPs is an acronym for ``weakly interactive massive particles''.
The best motivated candidate of this type is the lightest supersymmetric 
partner in supersymmetric extensions of the standard model of particle 
physics \cite{ehnos}. The mass range for which WIMPs provide the critical 
energy density for closing the universe is a few GeV to a few hundred GeV. 
The axion is a light pseudo-scalar particle whose existence has been 
postulated to explain why, in the context of the standard model of particle 
physics, the strong interactions conserve P and CP \cite{pq}. The likely
mass range for which axions provide the critical energy density for closing
the universe is $10^{-4}$ eV $< m_a < 10^{-7}$ eV \cite{arev}. 

Axions and WIMPs are the leading cold dark matter (CDM) candidates. Other
forms of dark matter are neutrinos and dark baryons. From the point of view
of galaxy formation, the defining properties of CDM are:
\begin{enumerate}
\item that CDM particles, unlike baryons, are guaranteed to interact with
their surroundings only through gravity, and
\item that CDM particles, unlike neutrinos, have negligibly small primordial
velocity dispersion.
\end{enumerate}
Studies of large scale structure formation support the view that the dominant 
fraction of dark matter is CDM. Moreover, if some fraction of the dark matter
is CDM, it necessarily contributes to galactic halos by falling into the
gravitational wells of galaxies and hence is susceptible to direct detection
on Earth. WIMPs are being searched for by looking for WIMP + nucleus elastic 
scattering in a laboratory detector \cite{psd88}. The nuclear recoil can be 
put into evidence by low temperature calorimetry, by ionization detection or 
by the detection of ballistic phonons.  Axions are being searched for by
stimulating their conversion to photons in a laboratory magnetic field 
\cite{ps83,llnl}. The experimental apparatus involves an electromagnetic 
cavity placed in the bore of a superconducting solenoid. When the resonant 
frequency of the lowest TM mode of the cavity equals the axion mass 
($h\nu=m_a \,c^2$), some galactic halo axions convert to microwave photons 
inside the cavity. If a signal is found in the cavity detector of dark 
matter axions, it will be possible to measure the energy spectrum of the
axions with great precision and resolution. (The energy resolution is 
limited only by the measurement integration time and the stability of the 
local oscillator with which the axion signal is compared.) Hence the 
question arises: what can be learned about our galaxy and the universe 
by analyzing such a signal?  The main purpose of this paper is to address 
this question by predicting properties of the CDM spectrum on Earth in 
terms of cosmological input parameters. Incidentally, all CDM candidates 
have the same phase space distribution, and hence the same spectrum on 
Earth, in the limit where their small primordial velocity dispersions 
are neglected.  We are motivated in part by the fact that knowledge of 
the spectrum may help in the discovery of a signal. 

In many past discussions of dark matter detection on Earth, it has been
assumed that the dark matter particles have an isothermal distribution, or 
an adiabatic deformation of an isothermal distribution. A strong argument 
in favor of this assumption is the fact that it predicts the rotation 
curve of the galaxy to be flat at large radial distances. Indeed, 
self-gravitating isothermal spheres always have density distributions 
$\rho(r)$ which fall off at large $r$ as $1/r^2$. Moreover, they have a 
``core radius'', i.e. a radius $a$ within which the density $\rho(r)$ 
no longer behaves as $1/r^2$ but goes to a constant as $r \rightarrow 0$. 
The behaviour may, for most practical purposes, be approximated by the 
function $\rho(r)=\rho(0)[1+(r/a)^2]^{-1}$. Thus, the contribution of 
an isothermal halo to the galactic rotation velocity goes to zero as 
$r \rightarrow 0$.  This feature of isothermal halos is attractive as 
well because it is known that, in spiral galaxies like our own, the 
rotation velocity at small radii may be entirely accounted for by 
the bulge and the disk.  In our galaxy, the core radius is such that 
roughly half of the rotation velocity squared at the solar radius $r_s 
\simeq 8.5$ kpc is due to the disk and bulge while the other half is due 
to the dark halo. Thermalization of the galactic halo has been argued 
to be the outcome of a period of ``violent relaxation'' \cite{lb} following 
the collapse of the protogalaxy. If it is strictly true that the velocity 
distribution of the dark matter particles is isothermal, then the only 
information that can be gained from its observation is the corresponding 
virial velocity and our own velocity relative to its standard of rest.

However, one may convince oneself that the velocity distribution of 
dark matter particles has a non-thermal component. Consider the fact 
that our closest neighbor on the galactic scale, the galaxy M31 in 
Andromeda, at a distance of order 730 kpc from us, is falling towards
our galaxy with a line-of-sight velocity of 120 km/sec. This motion
can be understood to be due to the mutual gravitational attraction between
the two galaxies: first they were receding from each other as part of the 
general Hubble flow, this relative motion was halted and now they are 
falling towards one another. We may use M31 as an indicator of the 
motion of any matter in our neighborhood. Moreover, if cold dark matter 
exists, then there is cold dark matter at every physical point in 
space (including everywhere we see nothing and which appears empty), 
because by Liouville's theorem the 3-dim. sheet in 6-dim. phase-space 
on which the CDM particles lie can not be ruptured. The thickness of that 
sheet is the tiny primordial velocity dispersion of the CDM particles, of 
order $10^{-12}$ for WIMPs and $10^{-17}$ for axions ($c=1$). The 
implication of the above is that, if CDM exists, there are CDM particles 
falling onto our galaxy continuously and from all directions. The motion 
of these particles gets randomized by gravitational scattering off giant 
molecular clouds, globular clusters and other inhomogeneities but complete 
thermalization of their velocity distribution occurs only after they 
have fallen in and out of the galaxy many times. As a result there 
are peaks in the velocity distribution of CDM particles at any physical 
point in the galaxy \cite{is92}. One peak is due to particles falling 
onto the Galaxy for the first time, one peak is due to particles falling 
out of the Galaxy for the first time, one peak is due to particles falling 
in for the second time, and so on. In particular, this is true on Earth. 
The width of the first two peaks, which we label n=1, is related to and 
is of order the velocity dispersion of the particles before they fall in 
for the first time. The width of the next two peaks (n=2) is expected to 
be somewhat larger as a result of scattering of the particles off 
inhomogeneities in the galaxy. The width of the next two peaks (n=3) 
is larger still because these particles have been scattered more. And so on.

One of the main purposes of this paper is to obtain estimates of the sizes
and velocity magnitudes of the velocity peaks on Earth. By ``size'', we mean 
the contribution of the peak to the local mass density of the halo. By 
``velocity magnitude'' , we mean the magnitude of the velocity vector of the  
particles in the peak as measured in the rest frame of the Galaxy, 
i.e. in a frame which is not corotating with the disk. The tool we 
use is the secondary infall model \cite{g77} of galactic halo 
formation. This model assumes a single overdensity in an expanding
universe. A halo forms around the overdensity because dark matter keeps 
falling onto it. The dark matter is assumed to have gravitational 
interactions only and to have zero initial velocity dispersion. The model 
also assumes spherical symmetry. Finally, in its original form, it assumes 
that the dark matter particles have zero angular momentum with respect to 
the center and hence that their motion is purely radial. Much progress 
\cite{fg84,eb85} in the analysis of the model was made as a result of 
the realization that the time-evolution is self-similar provided $\Omega =1$ 
and provided the initial overdensity has a special scale-free form; see 
Eq.(\ref{inm}). The parameter $\epsilon$ that appears in this ansatz is
related to the slope of the power spectrum of primordial density perturbations
at the galactic scale. Self-similarity means that the halo is time-independent
after all distances are rescaled by an overall time-dependent size $R(t)$ 
and all masses by a time-dependent mass $M(t)$. The self-similar solutions 
can be obtained numerically with great precision and some of their properties 
may be derived analytically.  When the parameter $\epsilon$ is in the range 
$0\le \epsilon \le 2/3$, the density profile $\rho(r) \sim 1/r^2$ and 
thus the rotation curve is flat.

However, for the purpose of estimating the sizes of velocity peaks, the 
secondary infall model without angular momentum is rather inadequate.  In 
particular, it tends to overestimate the size of the peaks due to dark matter 
particles falling in and out of the galaxy for the first time. Indeed, angular
momentum has the effect of keeping infalling dark matter away from the 
galactic center and this effect is largest for particles falling into the 
galaxy last. On the other hand, the presence of angular momentum destroys 
spherical symmetry and thus makes the actual evolution far more complicated 
and untractable. However, as will be explained in detail below, it is possible 
to include {\it the effect} of angular momentum into the secondary infall 
model without destroying its spherical symmetry by averaging over all possible
orientations of an actual physical halo \cite{stw95}. Moreover, the time
evolution of the model with angular momentum thus included is still 
self-similar provided the angular momentum distribution is of a particular 
scale-free form.  It was also found \cite{stw95} that angular momentum has 
the effect of making the halo contribution to the galactic rotation curve
go to zero at the galactic center, thus introducing an effective core radius 
for the halo mass distribution. We define the effective core radius $b$ to be 
the radius at which the halo contributes half of the galaxy's rotation 
velocity squared. For our Galaxy, $b$ is of order our own distance to the 
Galactic center.  This by itself suggests that the effect of angular 
momentum on the velocity peaks on Earth is not small. The model with 
angular momentum can be accurately solved on a computer. Its predictions 
for the effective core radius $b$, the local halo density, and the expected 
sizes and velocity magnitudes of the first few velocity peaks are tabulated
below for representative values of the input parameters, which are the age of 
the Universe, the parameter $\epsilon$ and the average amount of angular 
momentum. We also give an analytical treatment of the model under 
simplifying but realistic assumptions. It yields general formulae which 
may be used to estimate the expected sizes and the velocity magnitudes of the
velocity peaks for a wide range of the input parameter values.

In Section II we review the arguments of ref \cite{is92} why velocity peaks
in the cold dark matter spectrum on Earth are expected, and add some comments
of our own. In Section III we give a detailed description of the self-similar
infall model, without and with angular momentum. In Section IV we describe
how some of the model parameters are determined in terms of observed 
properties of our Galaxy and we give the results of the numerical 
integration of the model. Section V contains our analytical treatment of 
the model. Section VI summarizes our results.

\section{Phase space structure of cold dark matter halos.}
\label{sec:pss}

\begin{figure}
\psfig{file=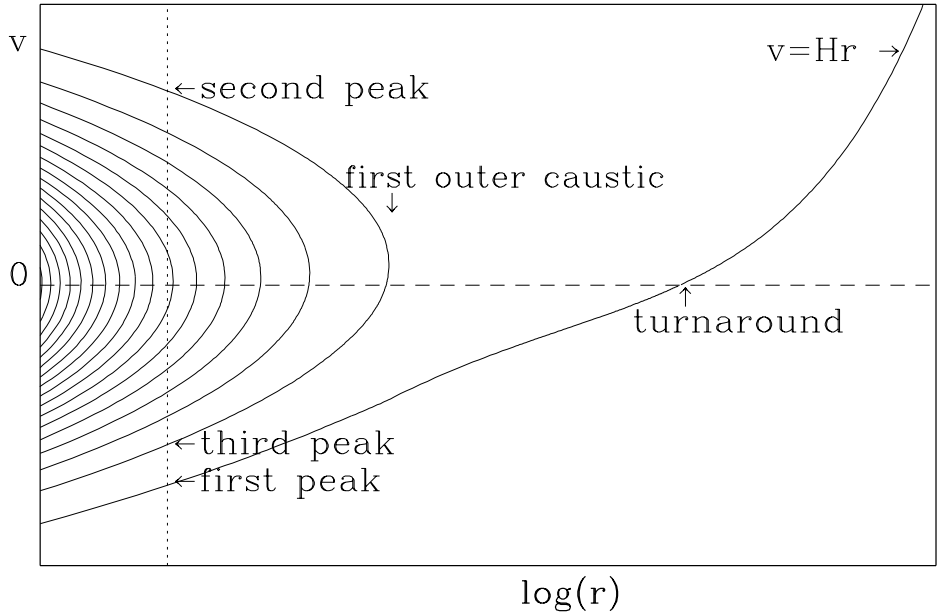,height=3.6in,width=5.6in}
\caption{Phase space distribution of the halo dark matter particles at a
fixed moment of time. The solid lines represent occupied phase-space cells.  
The dotted line corresponds to the observer position.  Each intersection of 
the solid and dotted lines produces a velocity peak.}
\label{fig:phsi1}
\end{figure}

In cold dark matter scenarios, the initial phase space
distribution is a very thin sheet near $\vec{v}=H\vec{r}$, 
where $H$ is the Hubble rate and $\vec{r}$ is the position 
relative to an arbitrarily chosen reference point.  The deviations 
from perfect Hubble flow which are present are associated with 
the primordial density perturbations that will produce galaxies and 
large scale structures by gravitational instability.  Where a galaxy 
(or some other object) forms and grows, the phase space sheet is 
folding itself up.  The process is illustrated in Fig. \ref{fig:phsi1} 
for the simplified case where a single spherically symmetric overdensity 
is present in an otherwise homogeneous universe and where all dark 
matter particles move on radial orbits through the center of 
the overdensity.  The line in the figure indicates the location of 
the dark matter particles in $(r,\dot{r})$ phase space at an 
instant of time.  $r$ is the distance to center of the overdensity and 
$\dot{r}= dr/dt$.  As time goes on the line ``winds up'' in the clockwise
direction, rotating most rapidly at the center.

Fig. \ref{fig:phsi1} shows that the velocity spectrum of cold dark 
matter particles on Earth, or anywhere else in the galaxy, has a series 
of peaks. One peak is due to particles falling onto the galaxy for 
the first time, passing by Earth while going towards the galactic center. 
A second peak is due to particles which are falling out of the galaxy 
for the first time, passing by Earth while going away from the 
galactic center.  A third peak is due to particles falling 
onto the galaxy for the second time.   A fourth peak is due 
to particles falling out of the galaxy for the second time.  And 
so on.  A rough estimate of the number $N$ of velocity peaks on 
Earth in this idealized case may be obtained as the ratio of 
the age of the galaxy $(\sim 10^{10}\, $ years) to the time 
$(\sim 0.5\times 10^8\,$ years) it takes a particle to fall 
to the center of the galaxy starting from rest at the Earth's 
location, with the result $N \sim 200$.  However, the presence of 
angular momentum of the dark matter particles tends to decrease $N$ 
by restricting the range of radii over which dark matter orbits vary. 
(In the extreme limit of circular orbits, $N = 1$.)  As will be 
seen below, this expectation \cite {is92} is confirmed by our calculations.
We shall also find that the number of peaks depends upon $\epsilon$.

Of course the description of a galactic halo presented in 
Fig. \ref{fig:phsi1} is much simplified.  In the remainder of 
this section, we discuss the sensitivity of the conclusion, 
that there are peaks in the cold dark matter velocity 
distribution on Earth, to the simplifying assumptions that 
were made.  In particular, we inquire into the effect of
\begin{enumerate}
\item the gravitational scattering of the dark matter particles by 
   inhomogeneities in the galaxy
\item the angular momentum that the dark matter particles 
   have with respect to the galactic center
\item the velocity dispersion that the dark matter particles 
   have before they fall onto the galaxy.
\end{enumerate}

\subsection{Scattering by inhomogeneities in the galaxy.}

The effect of the gravitational scattering of the dark matter 
particles by the inhomogeneities in the galaxy, such as stars, 
globular clusters and large molecular clouds, is to ``fuzz up'' 
the phase space sheets.  Consider a phase space sheet which 
in the absence of scattering produces on Earth a stream with 
unique velocity $\vec{v}$. The collective effect of scattering 
by a class of objects of mass $M$ and density $n$ is to diffuse 
the velocities in the stream over a cone of opening angle 
$\Delta \theta$ given by \cite{is92}
\begin{eqnarray}
(\Delta \theta)^2 &=& \int dt\int_{b_{min}}^{b_{max}}
{4G^2M^2\over b^2v^4}nv\, 2\pi bdb \nonumber \\
&\simeq& 2\times 10^{-7}\left( {M\over M_\odot}\right)^2 \ln 
\left( {b_{max}\over b_{min}}\right) \int dt\, 
{n \over v^3}  {(300\, {\rm km/s})^3\over 10^{10}\, 
{\rm year \,\, pc^{-3}} } \,\, ,
\label{dth}
\end{eqnarray}
where the time integral is over the past history of the particles 
in the stream, $b$ is the impact parameter of a scattering and 
$v$ is the velocity of the sheet relative to the scattering center. 
In the galactic disk, the giant molecular clouds are most likely 
the main contributors. With $M\sim 10^6\, M_\odot, \,\, n\sim 3\, 
{\rm kpc}^{-3}, \,\, b_{max} \sim$ kpc and $b_{min} \sim 20\, $ pc, 
they yield $\Delta \theta \simeq 0.05$ for dark matter
particles that have spent most of their past in the galactic disk.
The contributions due to globular clusters ($M \sim 5 
\times 10^5 M_\odot,\,\, n\sim 0.3 \, {\rm kpc}^{-3}$) and stars 
($M \sim M_\odot,\,\, n\sim 0.1\, {\rm pc}^{-3}$) are less important. 
At any rate, peaks due to dark matter particles that have spent 
much of their past in the central parts of the Galaxy are
likely to be washed out. On the other hand, the peaks due to 
dark matter particles which have fallen in and out of the Galaxy 
only a small number of times in the past are not erased by scattering. 

\subsection{Angular momentum.}

The presence of the rotating galactic disk clearly indicates that 
the baryons in our galaxy carry angular momentum. This angular 
momentum is thought to have been produced by the gravitational 
forces of nearby galaxies when ours started to form. 
One should expect the dark matter in the galactic halo to 
have similar amounts of angular momentum and hence to move 
on non-radial orbits. If an infalling particle's angular 
momentum is large enough, its distance of closest approach 
to the galactic center is larger than our own distance 
($\simeq$ 8.5 \,\, kpc) to the galactic center and hence it can 
not possibly reach us. It is nonetheless true that particles falling 
onto the galaxy for the first time reach us at all times, even if 
the typical distance of closest approach of such particles 
to the galactic center is much larger than 8.5 kpc. 

Indeed, consider all particles that reach their turnaround radius 
at a given time. ``Turnaround'' refers to the moment in a particle's 
history when it reaches zero radial velocity with respect to the 
galactic center for the first time, after its initial Hubble flow 
velocity has been halted by the gravitational pull of the galaxy 
and before it starts to fall onto the galaxy for the first time; 
see Fig. \ref{fig:phsi1}.  All particles that reach their turnaround 
radius at a given time are on a surface which, from a topological 
viewpoint, is a sphere enclosing the galactic center.  Let us call 
this surface the ``turnaround sphere''. By consulting a catalog of 
the galaxies in our neighborhood and plotting their radial velocities 
as a function of distance, one concludes that the radius of the present 
turnaround sphere is of order 1-2 Mpc for our galaxy. Consider the 
turnaround sphere at an arbitrary time t. At any point on that surface 
the angular momentum vector has a unique value parallel to the surface.  
Now, it is well known that a continuous vector field on a 2-sphere can 
not everywhere differ from zero. It must have at least two zeros. 

\begin{figure}
\psfig{file=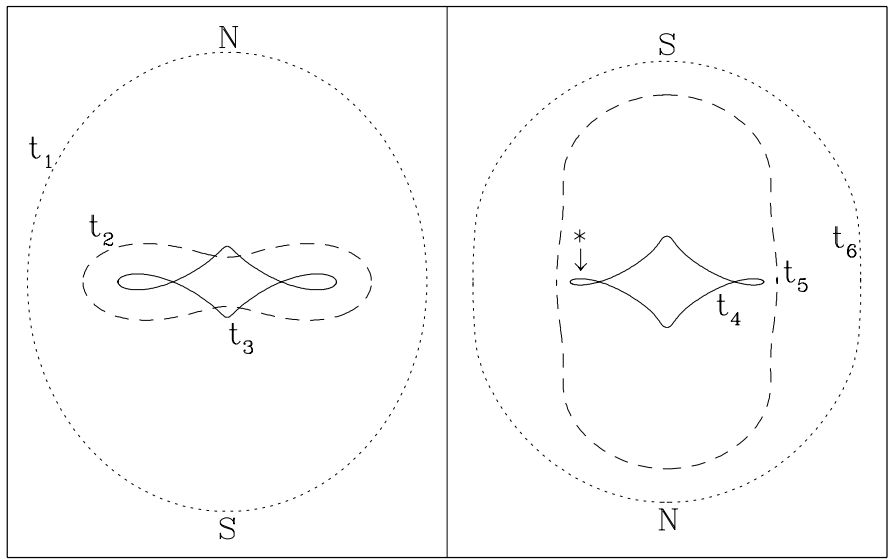,height=3.6in,width=5.6in}
\caption{Positions in physical space at successive moments in time
$t_1 < t_2 < \dots  <t_6$  of the particles on a turnaround sphere
that is intially rotating rigidly about the vertical axis.  The *
indicates the appearance of an inner caustic ring.}   
\label{fig:ts}
\end{figure}

Hence there are two places on any turnaround sphere where the angular 
momentum vanishes. The particles from these two locations will pass 
through the galactic center when they fall onto the galaxy next, 
producing two velocity peaks there.  By continuity (the phase space 
sheet can not tear), other particles on the turn-around sphere will 
produce two velocity peaks at any point sufficiently close
to the center.  Fig. \ref{fig:ts} shows the time evolution 
of a turn-around sphere which is initially rigidly rotating 
about an axis and which subsequently is moving under the 
influence of the gravitational potential of an isothermal sphere.
The figure demonstrates that any turnaround sphere passes (at least) 
twice by any point inside of it, once on the way in and once on the 
way out, assuming only that the point is inside the sphere both 
at its first and its second turnaround. By definition, second 
turnaround is when the sphere reaches its maximum size for the 
second time in its history, just after its first oscillation. Thus 
we find that when angular momentum is included, there are still two 
(possibly more but necessarily an even number) velocity peaks on Earth 
due to particles falling through the galaxy for the first time, two 
peaks or more due to particles falling through the galaxy for the 
second time, and so on. As we saw in the preceding subsection, these 
peaks are not erased by scattering off stars, globular clusters or 
giant molecular clouds. The sizes and velocity magnitudes of these 
peaks constitute the main topic of this paper.

Finally, let us note the effect angular momentum has on the 
caustics of a halo.  A ``caustic'' is a place where the dark 
matter density is large because the phase space sheet folds back 
there.  The density actually diverges at the caustic in the limit 
where the thickness of the phase space sheet goes to zero.  There 
is an outer caustic surface near the $n$-th turnaround radius with 
$n=2,3,4 ...$  ; see Fig. \ref{fig:phsi1}.  It can be shown that 
near a caustic surface, the dark matter density behaves as 
$\rho \sim \Theta(x)/\sqrt{x}$ + constant, where $x$ is the 
distance to the caustic surface and $\Theta(x)$ is the Heaviside 
function: $\Theta(x) =1$ for $x>0$ and $\Theta(x) =0$ otherwise.
Now, when angular momentum is absent, the center of the galaxy is 
a very special point because all dark matter particles go through 
the center at each orbital oscillation.  The dark matter density 
goes as $\rho \sim 1/r^2$ at the center if there is no angular 
momentum and the rotation curve is flat.  Thus, in the absence 
of angular momentum the galactic center is  a caustic point. When 
angular momentum is present, that caustic point spreads into a set 
of inner caustic rings.  Fig.\ref{fig:ts} shows the appearance of such 
an inner caustic ring for the case of axial symmetry. The fact that 
the caustic appears has nothing to do with axial symmetry however. 
Rather, it is a consequence of the fact that when a sphere is turned 
``inside out'', as illustrated in Fig.\ref{fig:ts}, a ring singularity
must appear on the surface at some point during the process.  Generally, 
the caustic ring appears near the place where the particles with the 
most angular momentum on a given turn-around sphere turn back at 
their distance of closest approach to the galactic center.  Outer 
caustics at the Earth's location are likely to be very much degraded
by scattering of the dark matter particles off inhomogeneities in the 
galaxy.  However, inner caustics associated with particles which have 
gone through the central parts of the galaxy only a small number of 
times in the past are not much degraded. The dark matter density on Earth 
could be much enhanced if we happen to be close to an inner caustic.
   
\subsection{On the velocity dispersion of infalling cold dark matter.}

In this study, when obtaining estimates of the average sizes 
and of the velocity magnitudes of the velocity peaks, we neglect 
the velocity dispersion $\delta v_{in}$ the cold dark matter has 
when it falls onto the galaxy for the first time. Presumably, this is 
a valid approximation provided $\delta v_{in}$ is much smaller than the 
velocity dispersion $\delta v_{gal}\sim 10^{-3}$ of the galaxy as a whole.  
We argue in this section that this condition is probably satisfied although 
we will not attempt to provide a reliable estimate for the size of 
$\delta v_{in}$.  The width $\delta v_n$ of the velocity peaks due 
to particles falling in and out of the galaxy for the $n$-th time, where 
$n$ is sufficiently small that the broadening effect of scattering of the 
particles by the galaxy's inhomogeneities can be neglected, is related 
to $\delta v_{in}$ by Liouville's theorem: 
$\delta v_n = \delta v_{in}(t_{*,n})[\rho_n / \rho(t_{*,n})]^{1/3}$ 
where $\rho_n$ is the contribution to the local halo density from 
particles in the $n$-th peak, and $\delta v_{in}(t_{*,n})$ and 
$\rho(t_{*,n})$ are the velocity dispersion and density those 
particles had at the time $t_{*,n}$ of their first turn-around. 

	Let us emphasize that the values of the peak widths
$\delta v_n$ may some day be measured in a direct CDM detection
experiment and that such data would provide information about 
our universe which is not readily accessible by other 
means.  Also, if the widths of some peaks are small enough
the sensitivity of the cavity detector of dark matter axions 
is improved by looking for narrow peaks.  In the case of the 
present LLNL experiment \cite{llnl}, which does look for narrow peaks
in addition to looking for a signal whose width is set by the 
galaxy's overall velocity dispersion $\delta v_{gal}\sim 10^{-3}$, 
the sensitivity of the search is improved if there is a velocity 
peak with $\delta v_n$ less than about $10^{-8}$ and with a 
fraction of the local density larger than about 1\%.

	Turning to the question how large $\delta v_{in}$ may be,
let us start by describing the {\it primordial} velocity dispersion 
which is the contribution that is present even if the universe were 
completely homogeneous, i.e., it is the value of $\delta v_{in}$ 
if our galaxy were the only density perturbation in the universe. For 
WIMPs, the primordial velocity dispersiom  $\delta v_W$ is due to the 
finite temperature $T_D$ the WIMPs have when their kinetic energies 
decouple from the primordial heat bath.  Thus 
$\delta v_W \sim (2T_D/m)^{1/2}(R_D/R_0)$, where $m$ is the WIMP mass, 
and $R_D$ and $R_0$ are the scale factors at temperature $T_D$ and now.  
For $m\sim 50\, {\rm GeV}$ and $T_D \sim 1\, {\rm MeV}$, one has 
$v_W \sim 10^{-12}$ which is very small.  For axions, the primordial 
velocity dispersion is due to the inhomogeneity of the axion field 
at temperature $T_1 \simeq$ 1 GeV and time $t_1 \simeq 2\times 10^{-7}$ 
sec when the axion mass becomes equal to the Hubble expansion rate.  If 
there is no inflation after the Peccei-Quinn phase transition at
which the $U_{PQ}(1)$ symmetry gets spontaneously broken, then
the scale of inhomogeneity of the axion field is of order the 
horizon scale at time $t_1$ and the primordial axion velocity 
dispersion today is therefore $\delta v_a \sim 
(m_a t_1)^{-1} (R_1/R_0) \sim 10^{-17} \times (10^{-5} \, eV/m_a)$. 
If there is inflation after the Peccei-Quinn phase transition, then 
the axion field gets homogenized over enormous distances and 
$\delta v_a$ is exponentially small.   

\begin{figure}
\psfig{file=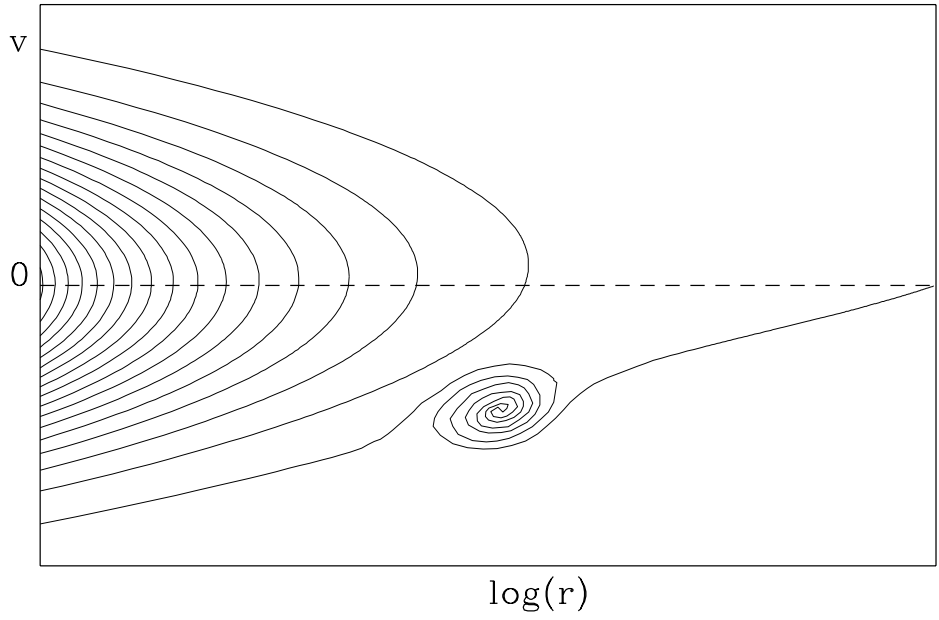,height=3.6in,width=5.6in}
\caption{A small scale sub-clump falling into the galaxy for the first time.}
\label{fig:phsi2}
\end{figure}

	The primordial velocity dispersion, $\delta v_W$ or 
$\delta v_a$, discussed in the preceding paragraph is the
thickness of the CDM phase space sheet.  It constitutes
a lower bound on the velocity dispersion $\delta v_{in}$ of 
infalling CDM.  Additional velocity dispersion is expected 
because the phase space sheet may wrap itself up on smaller 
scales than that of the galaxy as a whole, as illustrated in 
Fig.\ref{fig:phsi2}.  The phase space sheet wraps itself up 
wherever an overdensity has grown by gravitational instability 
past the linear regime ($\delta \rho / {\rho} < 1$) into the 
non-linear one ($\delta \rho / {\rho} > 1$).  In theories of 
structure formation based upon cold dark matter, the spectrum of 
primordial density perturbations is flat, i.e., it has approximately 
equal power on all length scales.  The matter density perturbations 
do not grow till the time $t_{eq}$ of equality between the matter 
and radiation energy densities.  After $t_{eq}$, all density 
perturbations which have wavelength less than the horizon (this 
includes all length scales of order a galaxy size) grow together 
at the same rate and therefore they all reach the non-linear
regime at approximately the same time.  In the standard CDM cosmology,
the smaller scale clumps reach the non-linear regime somewhat earlier 
because the processed spectrum of density perturbations is not exactly 
flat on galaxy scales but is slightly tilted with more power on small 
scales. 

What happens in the non-linear regime is far from obvious.  The 
rate of growth of an overdensity in the non-linear regime is of order 
$\sqrt{G \rho}$, where $\rho$ is its mean density. Indeed $1/\sqrt{G \rho}$
is of order the free infall time, which is also the time necessary to 
produce a new fold in the phase space sheet.  At the start of the 
non-linear regime, as we just argued, all overdensities have densities 
of the same order of magnitude and they therefore grow at comparable 
rates by locally wrapping up the phase space sheet.  However, 
overdensities of large physical size will tidally disrupt and 
therfore inhibit the growth of overdensities of smaller physical 
size in their neighborhood.  In the immediate vicinity of our galaxy, 
there are no visible overdensities other than nearby dwarf galaxies 
such as the Magellanic clouds.  Dark matter and accompanying baryons 
are nonetheless falling onto the galaxy now for the first time. It 
is the velocity dispersion of this unseen matter that we are interested 
in.  If this matter is in large clumps, one might expect it to light up 
stars and thus become visible. On the other hand, it could be in clumps 
which have not lit up for some reason.  However, any known object smaller 
than a galaxy (e.g. stars, globular clusters, large molecular clouds, dwarf 
galaxies) has velocity dispersion smaller than $\delta v_{gal}\simeq 10^{-3}$, 
and dark matter objects should be expected to be less clumped than baryonic 
objects because they can not dissipate their energy.  On this basis, it 
seems safe to assume that $\delta v_{in}$ is considerably less than 
$\delta v_{gal}$.    

	There is a particular kind of clumpiness which is expected to
affect axion dark matter if there is no inflation after the Peccei-Quinn 
phase transition.  This is due to the fact that cold dark matter 
axions are inhomogeneous with $\delta \rho / \rho \sim 1$ over 
the horizon scale at temperature $T_1 \simeq$ 1 GeV when they 
are produced at the start of the QCD phase-transition, combined 
with the fact that their velocities are so small that they do not 
erase these inhomogeneities by free-streaming before the time $t_{eq}$ 
when matter perturbations can start to grow.  These particular 
inhomogeneities in the axion dark matter are immediately in the non-linear 
regime after time $t_{eq}$ and thus form clumps, called ``axion 
mini-clusters'' \cite{hr88,kt93,kt96}.  These have \cite{kt96} mass 
$M_{mc} \simeq 10^{-13} M_\odot$ and size $l_{mc} \simeq 10^{12}$ cm, and 
therefore their associated velocity dispersion  
$v_{mc}= \sqrt{GM_{mc}/l_{mc}} \simeq 10^{-10}$ at time $t_{eq}$.  This 
velocity dispersion increases by about a factor 10 from $t_{eq}$ till the 
onset of galaxy formation because of the hierarchical clustering of the 
axion mini-clusters.  This yields $v_{mc} \simeq 10^{-9}$ as the contribution 
of mini-clusters to the velocity dispersion $\delta v_{in}$ of infalling 
axions if there is no inflation after the Peccei-Quinn transition.

\section{Self-similar infall models.}
\label{sec:ssi}

\subsection{The radial infall model.}
\label{sec:ri}

The tool we use to obtain estimates of the sizes and the velocity 
magnitudes of the highest energy peaks is the secondary infall 
model of galactic halo formation.  In its original form, this 
model is based on the following assumptions:
\begin{enumerate}
\item the dark matter is non-dissipative
\item it has negligible initial velocity dispersion
\item the gravitational potential of the galaxy is spherically 
symmetric and is dominated by the dark matter contribution
\item the dark matter particles move on radial orbits through 
the galactic center.
\end{enumerate}
Assumption 1 means that the only force acting upon the dark matter 
particles is the gravitational pull of the galaxy. Assumption 2 
states that before the galaxy starts to form, at some initial time 
$t_i$, all the dark matter particles at the same position $\vec{r}_i$ 
relative to the galactic center move with the same velocity $\vec{v}_i$.  
Provided $t_i$ is chosen early enough, this initial velocity is given by 
the Hubble expansion:
\begin{equation}
\vec{v}_i=H(t_i)\vec{r}_i \,\, .
\label{hf}
\end{equation}
$H(t_i)$ is the Hubble rate at time $t_i$.  The issue of the validity 
of assumption 2 is discussed at length in the previous section.  
Henceforth, we will take its validity for granted.  Assumption 3 is 
realistic because the gravitational potential of a galaxy as a whole 
(luminous plus dark matter) is nearly spherically symmetric even if 
the distribution of its luminous matter is not spherically symmetric 
at all. Assumption 4 is the most doubtful. As was discussed in the 
previous section, the rotating disks of spiral galaxies show that their 
baryons carry angular momentum and one should expect the dark matter to 
have similar amounts of angular momentum and hence to move on non-radial 
orbits with distances of closest approach to the galactic center at least 
of order the radius ($\sim 10$ kpc) of the disk.  Assumption 4 is motivated 
mainly by simplicity.  Below, in the next subsection, we will generalize the 
model to rid it of this assumption.  For clarity, we refer to the model 
with the fourth assumption included as the radial infall model.

Let us call $M_i$ the mass inside $r_i$ at the initial time $t_i$. 
In a perfectly homogeneous and flat universe, $M_i$ is equal to
%\begin{eqnarray}
\begin{equation}
M_i^{\Omega =1} = {4\pi \over 3} \rho(t_i)r_i^3
%\nonumber \\
= {r_i^3 H(t_i)^2 \over 2G} ={2r_i^3 \over 9Gt_i^2} \, ,
\label{mi}
%\end{eqnarray}
\end{equation}
if we take $t_i$ to be in the matter dominated epoch. Instead,
\begin{equation}
M_i(r_i)={2r_i^3 \over 9Gt_i^2} +\delta M_i(r_i) \, .
\label{mi2}
\end{equation}
where $\delta M_i(r_i)$ is a spherically symmetric overdensity. The dark 
matter shell which is initially at radius $r_i$ has initially the radial 
velocity
\begin{equation}
v_i(r_i)=H(t_i)r_i=2r_i/3t_i \,\, ,
\label{vi}
\end{equation}
assuming that $t_i$ is small enough so that the very earliest
deviations from perfect Hubble flow may be neglected. The position 
$r(r_i,t)$ of each shell at time $t$ is determined by solving 
the equations
\begin{equation}
{d^2r \over dt^2} =-{G\, M(r,t) \over r^2} \,\, ,
\label{d2r}
\end{equation}
\begin{equation}
M(r,t)=\int_0^{\infty}dr_i {dM_i \over dr_i} \Theta(r-r(r_i,t)) \,\, ,
\label{mrt}
\end{equation}
with the initial conditions given in Eq. (\ref{vi}). $\Theta(x)$ is 
the Heaviside function, defined earlier.

The qualitative evolution of the dark matter distribution in
phase-space $(r,\dot{r})$ may be described as follows. Initially,
the dark matter particles are located on the line $\dot{r} = H(t_i) r$. 
As time goes on, this line ``winds up'' in a clockwise fashion 
rotating most rapidly near $r=\dot{r}=0$.  Fig.\ref{fig:phsi1} shows  
the line on which the dark matter particles are located at a 
particular moment in time. 
  
The radius $r(r_i,t)$ of a given shell initially increases till it 
reaches a maximum value $r_*(r_i)$ at a time $t_*(r_i)$. $r_*(r_i)$ 
and $t_*(r_i)$ are called the turnaround radius and turnaround time 
of shell $r_i$. After $t_*(r_i)$, the radius of shell $r_i$ will
oscillate with decreasing amplitude.  As long as it does not cross 
any other shells, the mass interior to shell $r_i$ is constant, with 
value $M_i$, and its motion is the well-known motion of a particle 
attracted by a central mass $M_i$ in the limit of zero angular 
momentum.  Shell $r_i$ does not cross any other shells till some 
time after $t_*(r_i)$, when it is falling onto the galactic center 
for the first time. Thus, one readily finds:
\begin{mathletters}
\label{trta}
\begin{equation}
t_*(r_i) = {\pi \over 2} \sqrt{r_*(r_i)^3\over 2GM_i(r_i)} \,\, , \,\,\,\,\,
\text{and}
\label{trtaa}
\end{equation}
\begin{equation}
r_*(r_i) = r_i{M_i(r_i) \over \delta M_i(r_i)} \,\, .
\label{trtab}
\end{equation}
\end{mathletters}
After a given shell crosses other shells its motion depends on 
that of the other shells and becomes more difficult to determine. 

Much progress in the analysis of the model came about as a result of 
the realization \cite{fg84,eb85} that Eqs.(\ref{d2r}) and (\ref{mrt}) 
have self-similar solutions for appropriate initial conditions. 
A solution is self-similar if it remains identical to itself 
after all distances have been rescaled by a time-dependent 
length $R(t)$ and all masses by a time-dependent mass 
$M(t)$. $R(t)$ is taken to be the radius at which dark matter 
particles are turning around at time $t$; see Fig. \ref{fig:phsi1}. 
$M(t)$ is taken to be the mass interior to $R(t)$ at time $t$. 
So, $R(t)=r_*(r_i)$ and $M(t)=M_i(r_i)$ with $r_i$ such that 
$t_*(r_i)=t$.  A self-similar solution has the properties:
\begin{equation}
M(r,t)=M(t) {\cal M}(r/R(t)) \, ,
\label{mss}
\end{equation}
and
\begin{equation}
r(r_i,t)=r_*(r_i)\lambda(t/t_*(r_i)) \, ,
\label{rss}
\end{equation}
where ${\cal M}$ and $\lambda$ are functions of a single
variable. Let us verify that indeed the evolution is self-similar
for appropriate initial conditions. Substituting Eqs (\ref{mss}) and 
(\ref{rss}) into Eq. (\ref{d2r}) and using Eq. (\ref{trtaa}), one 
finds:
\begin{equation}
{d^2\lambda \over d\tau^2}=-{\pi^2 \over 8\lambda^2}{M(t) 
\over M_i(r_i)} {\cal M}\left( {r_*(r_i)\lambda \over R(t)}\right) \,\, ,
\label{d2l}
\end{equation}
where $\tau \equiv t/t_*(r_i)$. We want the RHS of Eq. (\ref{d2l}) 
to depend only upon $\tau$ and $\lambda (\tau )$. This happens for 
the initial condition:
\begin{equation}
{\delta M_i(r_i) \over M_i(r_i)} = \left({M_0 \over M_i(r_i)} 
\right)^\epsilon \,\, ,
\label{inm}
\end{equation}
where $M_0$ and $\epsilon$ are parameters. $\epsilon$ should be 
in the range $0 \leq \epsilon \leq 1$, since $\epsilon = 0$
corresponds to the extreme case of a $r_i$-independent overdensity 
whereas $\epsilon = 1$ corresponds to the extreme case of an excess 
point mass located at $r=0$.  The initial density profile (\ref{inm}) 
does not have any feature that would distinguish an epoch in the 
evolution of the galactic halo from other epochs. It is this 
``scale free'' property that makes the initial density profile (\ref{inm}) 
consistent with self-similarity, as we are about to show. From now on, 
for the sake of convenience and following Fillmore and Goldreich, we 
will use $M_i$ instead of $r_i$ to label the shells.  Using 
Eqs. (\ref{mi2}) and (\ref{inm}), and neglecting terms of order 
$\delta M_i/M_i$ versus terms of order one, we find that 
Eqs. (\ref{trta}) become
\begin{mathletters}
\label{trta2}
\begin{equation}
t_*(r_i) = {3\pi \over 4} t_i \left( {M_i \over M_0} \right) 
^{3\epsilon /2} \,\, , \label{trta2a}
\end{equation}
\begin{equation}
r_*(M_i) = \left[{8 \over \pi^2} t_*^2(M_i)GM_i\right] ^{1/3} \,\, .
\label{trta2b}
\end{equation}
\end{mathletters}
Hence
\begin{mathletters}
\label{trta3}
\begin{equation}
M(t) = M_0  \left( {4 t\over 3\pi t_i} \right) ^{2/3\epsilon} \,\, , \,\,\,\,\,
\text{and}
\label{trta3a}
\end{equation}
\begin{equation}
R(t) = \left[{8t^2G \over \pi^2} M(t)\right] ^{1/3} \,\, .
\label{trta3b}
\end{equation}
\end{mathletters}
Therefore
\begin{mathletters}
\label{trta4}
\begin{equation}
{M(t) \over M_i} = {M(t)\over M(t_*(M_i))}=\left( t\over 
t_*(M_i)\right)^{2/3\epsilon}=\tau^{2/3\epsilon} \,\, ,
\label{trta4a}
\end{equation}
\begin{equation}
{r_*(M_i)\over R(t)} = {R(t_*(M_i)) \over R(t)} =\left( {t_*(M_i) 
\over t}\right) ^{2/3 +2/9\epsilon} =\tau^{-2/3 -2/9\epsilon}\,\, .
\label{trta4b}
\end{equation}
\end{mathletters}
Thus, Eq. (\ref{d2l}) has the desired form:
\begin{equation}
{d^2\lambda \over d\tau^2}=-{\pi^2 \over 8}{\tau^{2/3\epsilon} \over 
\lambda^2} {\cal M}\left( {\lambda \over \tau^{2/3 +2/9\epsilon} }\right) \, .
\label{d2l2}
\end{equation}
Similarly, using Eq. (\ref{mss}) and Eqs. (\ref{trta4}), we rewrite 
Eq. (\ref{mrt}) as
\begin{eqnarray}
{\cal M}(\xi ) &=& {M(\xi R(t),t)\over M(t)}
= \int_0^{\infty}{dM_i \over M(t)}\, \Theta\left(\xi R(t)-r_*(M_i)
\lambda \left({t\over r_*(M_i)}\right)\right) \nonumber \\
&=& {2\over 3\epsilon}\int_1^{\infty}{d\tau \over \tau^{1+2/3\epsilon}}\,
\Theta\left(\xi -{\lambda (\tau ) \over \tau^{2/3+2/9\epsilon}}\right)\,\, ,
\label{mxi}
\end{eqnarray}
which also has the desired form. $\tau$ varies from 1 to $\infty$. 
The boundary conditions at $\tau =1$ are:
\begin{equation}
\lambda (1) =1, \,\,\,\,\,\, {d\lambda \over d\tau}(1)=0 \,\, .
\label{bc}
\end{equation}
Fillmore and Goldreich \cite{fg84} integrated Eqs. (\ref{d2l2}) 
and (\ref{mxi}) numerically for various values of $\epsilon$.
They also derived analytically the behaviour of ${\cal M} (\xi )$ 
when $\xi \rightarrow 0$ using adiabatic invariants to obtain
the motion of the shells.  Bertschinger \cite{eb85} analyzed the 
case $\epsilon =1$. 

Let us mention in passing that the ratio of the density at the turnaround 
radius to the critical density is  $\rho_*/\rho_c 
= 9\pi^2/[16(3\epsilon +1)]$. 

\subsection{Secondary infall with angular momentum.}
\label{sec:am}

As was emphasized earlier, the assumption that the infalling 
dark matter is devoid of angular momentum with respect to the 
galactic center is inadequate for the calculation of the velocity 
peaks which are the main topic of this paper.  However, it is 
possible to include the effect of angular momentum into the 
secondary infall model while keeping the model tractable.  We 
will do this in two steps.  First we will assign the same value 
of angular momentum magnitude to all particles in a given shell.  
Second, we will assign the particles in a given shell a distribution 
of angular momentum magnitudes.  The model which results from the 
first step is not so realistic but it is easier to explain.

\subsubsection{Single magnitude of angular momentum for all 
particles on a shell.}

Let's assume first that the particles belonging to shell $M_i$ all have 
the same {\it magnitude} of angular momentum $l(M_i)$ and that they
all have, at some initial time, the same radial coordinate $r(M_i,t)$
and the same radial velocity $v_r(M_i,t)= \partial r(M_i,t)/\partial t$.
We further assume that at any point $\vec{r}(M_i,t)=\hat{r} r(M_i,t)$ 
on shell $M_i$, all particles in the shell have their transverse 
velocities $\vec{v}_\perp (M_i,\hat{r},t) =l(M_i) \hat{\varphi}/r(M_i,t)$ 
isotropically distributed about $\hat{r}$, i.~e. each direction
$\hat{\varphi}$ perpendicular to $\hat{r}$ is equally much represented. 
As a result of these assumptions, each shell remains spherical as it 
moves through the spherically symmetric mass distribution $M(r,t)$ due 
to all other shells. Initially, at time $t_i$, the shell $M_i$ has radial 
velocity
\begin{equation}
v_r(M_i,t)=H(t_i)r(M_i,t_i)= {2 r(M_i,t_i) \over 3 t_i} \,\, .
\label{vri}
\end{equation}
Eqs. (\ref{mi}) and (\ref{mi2}) hold as before. Provided
$l(M_i)$ is not too large, the turnaround radius $r_*(M_i)$ and 
time $t_*(M_i)$ are still given by Eqs. (\ref{trta}) to a very 
good approximation. We will use these equations, neglecting the 
corrections therein due to  $l(M_i) \neq 0$ and find that 
self-similarity is possible. Note, however, that if the 
corrections to Eqs. (\ref{trta}) due to $l(M_i) \neq 0$ are 
included one still finds self-similarity to be possible.
The reason we neglect the corrections is not to obtain 
self-similarity but because these corrections are truly small 
for realistic values of angular momentum.

To obtain self-similarity we assume as a necesary, but no longer 
sufficient, condition that the initial mass distribution is 
given by the scale free power law of Eq. (\ref{inm}). 
Eqs. (\ref{trta2}), (\ref{trta3}), (\ref{trta4}) are then still 
valid as well. Each dark matter particle satisfies:
\begin{equation}
{d^2r \over dt^2} ={l^2 \over r^3} -{GM(r,t) \over r^2}\,\, .
\label{eql}
\end{equation}
Substituting therein
\begin{mathletters}
\label{rm}
\begin{equation}
r(M_i,t)=r_*(M_i)\lambda \left({t\over t_*(M_i)}\right) \,\, ,
\label{rma}
\end{equation}
\begin{equation}
M(r,t)=M(t) {\cal M} \left({r\over R(t)}\right) \,\, ,
\label{rmb}
\end{equation}
\end{mathletters}
one obtains, using Eqs. (\ref{trta2}) - (\ref{trta4}):
\begin{equation}
{d^2\lambda \over d\tau^2}= {l(M_i)^2t_*(M_i)^2 \over r_*(M_i)^4
\lambda^3}-{\pi^2 \over 8}{\tau^{2/3\epsilon} \over \lambda^2} 
{\cal M}\left({\lambda \over \tau^{2/3 +2/9\epsilon} }\right) \,\, ,
\label{d2ll}
\end{equation}
where $\tau=t/t_*(M_i)$ as before. To obtain self-similar solutions, 
we must make the additional assumption that
\begin{equation}
l(M_i)=j{r_*(M_i)^2 \over t_*(M_i)} \,\, ,
\label{ssl}
\end{equation}
where $j$ is a constant. Then
\begin{equation}
{d^2\lambda \over d\tau^2}= {j^2 \over \lambda^3}-{\pi^2 \over
8}{\tau^{2/3\epsilon} \over \lambda^2} {\cal M}\left( 
{\lambda \over \tau^{2/3+2/9\epsilon} }\right) \,\, .
\label{d2ll2}
\end{equation}
Eq. (\ref{mxi}) for ${\cal M}(\xi )$ and the boundary conditions 
(\ref{bc}) remain unchanged.

\begin{figure}
\psfig{file=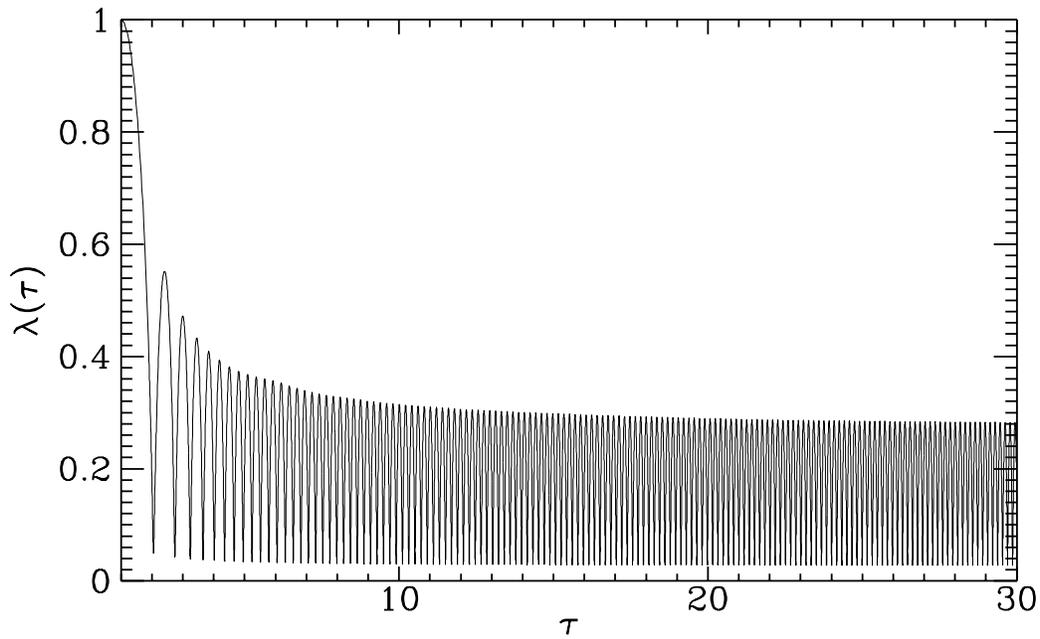,height=3.6in,width=5.6in}
\caption{The function $\lambda (\tau )$ for $\epsilon =0.2$, $j=0.2$.} 
\label{fig:ltau}
\end{figure}

The question arises how realistic the model is in the above
form.  Consider shell $M_i$ near its turnaround time $t_*(M_i)$. 
The actual angular momenta of the particles in the shell are 
of course not distributed as in the model.  The model assumes 
$\vec{v}_\perp$ to be isotropically distributed about $\vec{r}$. 
Instead, angular momentum has a unique value $\vec{l}(M_i,\vec{r})$ 
at each point, with $\vec{l}(M_i,\vec{r})$ changing from point to 
point on the shell. As was discussed in Section \ref{sec:pss},
$\vec{l}(M_i,\vec{r})$ must have at least two zeros on the shell. This 
implies that there are necessarily some particles in the shell which 
will pass through the center of the galaxy and, by continuity, other 
particles  will reach the Earth as well.  There are two velocity peaks 
in the spectrum of cold dark matter particles on Earth due to particles 
falling into and out of the galaxy for the first time, two peaks due 
particles falling into and out of the galaxy for the second time, etc.  
In contrast, in the above model, after a certain galactic age none of 
the particles falling onto the galaxy for the first time reach the 
Earth because these particles have too much angular momentum. Their 
distance of closest approach to the galactic center exceeds our own 
distance to the galactic center.  If we turn for a moment to the results
of the computer simulations, we see that for the example of 
Figs.\ref{fig:phs} and \ref{fig:ltau}, in which $j=0.2$, only particles 
which are falling in and out of the galaxy for the $n$-th time with 
$n>3$ can presently reach us.

\subsubsection{Distribution of magnitude of angular momenta on 
a shell.}

In reality, particles at different locations on a given shell have 
different values of vector angular momentum.  As a result, the time
evolution of a shell is not spherically symmetric when angular
momentum is present. This is illustrated by Fig.\ref{fig:ts} in a 
special axially symmetric case.  However, we can restore spherical 
symmetry by averaging over all possible orientations of a 
physical halo.  This corresponds to adopting the model of 
the previous subsection but with a distribution of magnitudes 
of angular momentum for the particles in each shell.  Let each 
shell $M_i$ have a fraction $n_k (M_i)$ of particles with magnitude 
of angular momentum $l_k(M_i)$ where $k= 1, \ldots , K$. To 
obtain self-similar solutions, $n_k(M_i)$ must be independent 
of $M_i$ and
\begin{equation}
l_k (M_i)=j_k {r_*(M_i)^2 \over t_*(M_i)} \,\, .
\label{lssm}
\end{equation}
The equations for self-similar solutions are then
\begin{mathletters}
\label{ml}
\begin{equation}
r_k (M_i,t)=r_*(M_i)\lambda_k \left({t\over t_*(M_i)}\right) \,\, ,
\label{mla}
\end{equation}
\begin{equation}
{d^2\lambda_k \over d\tau^2}= {j^2_k \over \lambda^3_k}-{\pi^2
\over 8}{\tau^{2/3\epsilon} \over \lambda^2_k } {\cal M}\left(
{\lambda_k \over \tau^{2/3 +2/9\epsilon} }\right) \,\, ,
\label{mlb}
\end{equation}
\begin{equation}
{\cal M}(\xi ) = {2\over 3\epsilon}\sum_{k=1}^K n_k \int_1^{\infty}{d\tau
\over \tau^{1+2/3\epsilon}}\, \Theta\left(\xi -{\lambda_k (\tau ) 
\over\tau^{2/3+2/9\epsilon}}\right)\,\, ,
\label{mlc}
\end{equation}
\begin{equation}
\sum_{k =1}^K n_k =1 \,\, .
\label{mld}
\end{equation}
\end{mathletters}
In our calculations, we shall take $j_k$ to be distributed according to 
the density
\begin{equation}
{dn \over dj}={2j \over j_0^2} \exp(-j^2/j_0^2)  \,\, .
\label{nj}
\end{equation}
Let us explain this choice, starting with the behaviour of 
$dn / dj$ near $j =0$.  The angular momentum field on a 
sphere must have at least two zeros. Let us choose the origin
($\theta =0$) of polar coordinates to be at one of them.  Assuming 
that the zero is simple, the Taylor expansion of the magnitude
of angular momentum function $j(\theta,\phi)$ in powers of 
$\theta$ starts with the term linear in $\theta$: $j \sim \theta$.
Then we have: $dn/dj \sim dn/d\theta \sim \theta \sim j$.  
The cutoff at large $j$ in Eq. (\ref{nj}) was chosen to be Gaussian
for the sake of convenience.  The actual distribution likely has 
a sharp cutoff, with a maximum value of angular momentum, but that 
is very similar to a Gaussian.  We express our results below in 
terms of the average over the distribution, $\bar j = \sqrt{\pi} j_0/2$.  

A benefit of including angular momentum into the secondary infall 
model is to produce galactic halos with an effective core radius.
The radial infall model, i.e. the model without angular momentum,
produces flat rotation curves for $0 < \epsilon < 2/3$.  Adding 
angular momentum has the effect of depleting the inner halo and 
hence of making the halo contribution to the rotation curve go 
to zero as $r \rightarrow 0$.  This is desirable because, in spiral 
galaxies like our own, it is the sum of the contributions from 
the halo, the disk and the bulge that produces flat rotation curves,
and the central parts of the galaxy are known to be dominated by the 
bulge and the disk.  We define the ``effective halo core radius'' $b$ 
as the radius at which half of the rotation velocity squared is due 
to the halo.  In our galaxy $b$ is estimated to be of order 10 kpc.  
We will find below that this implies $\bar j \sim 0.2$ in the model.

The secondary infall model with a distribution of angular momentum
described in this subsection still has shortcomings due to the fact 
that the model averages over all possible orientations of a physical 
halo.  In particular, the model is only able to produce estimates of 
the {\it average} sizes of velocity peaks.  The averages are over 
all locations at the same distance from the galactic center as we are.  
At some of these locations, a particular velocity peak may be much 
larger than average because that location happens to be close to an 
inner caustic. 

\subsection{Inclusion of baryons}
\label{sec:bi}
 
One may also wish to include the effect of the gravitational potential 
of the galactic bulge and disk.  As was already noted, the disk and bulge 
of our galaxy are conspiring with its dark matter halo to produce an 
everywhere approximately flat rotation curve.  We may reasonably assume 
that this was also true in the past because most spiral galaxies are 
observed to have approximately flat rotation curves and they do not all 
have the same age.  This suggests a simple way to include the effect of 
baryons in the self-similar secondary infall model, to wit:  first obtain 
for given $\epsilon$ the mass function $\cal M (\xi)$ of the model without 
angular momentum and then use that mass function and Eq.(\ref{mlb}) to 
obtain the phase space distribution of the dark matter in the model with 
angular momentum.  We will find below that including the gravitational 
field of the disk and bulge in this manner does not have much effect upon 
the sizes of the highest energy peaks but it does shift their kinetic 
energies upward by deepening the potential well at our location.    

\subsection{$\epsilon$ and the spectrum of initial density 
perturbations.}
\label{sec:eps?}

\begin{figure}
\psfig{file=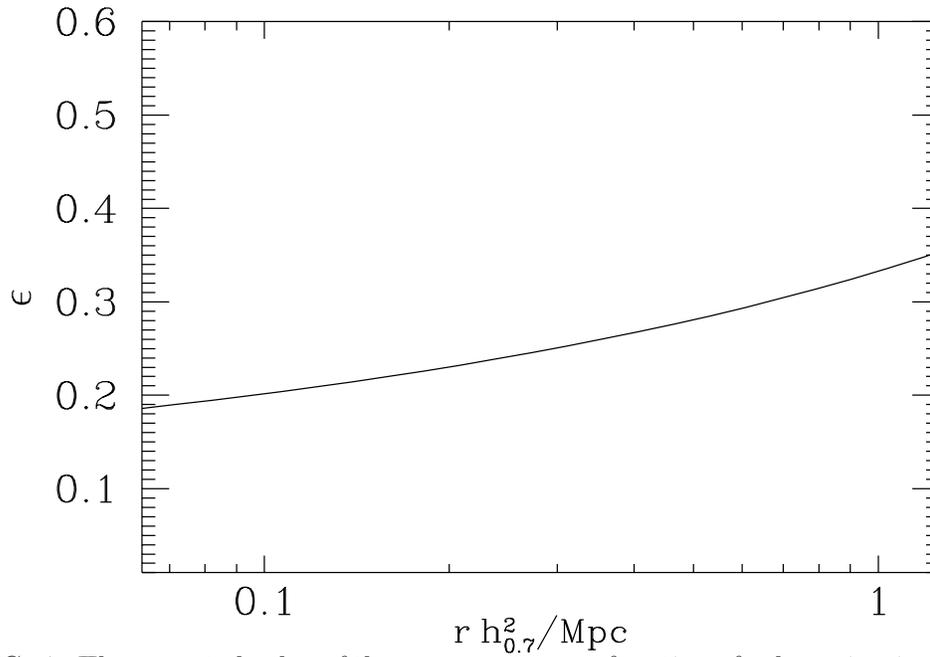,height=3.6in,width=5.6in}
\caption{The expected value of the $\epsilon$ parameter as a function of 
galaxy size, in models of structure formation based upon cold dark matter
and a flat (Harrison-Zel'dovich) spectrum of primordial density 
perturbations.  We defined $h_{0.7} \equiv h/0.7$}
\label{fig:ps}
\end{figure}

The spectrum of the cosmological density perturbations which give rise
to galaxies contains information about the likely value of the parameter
$\epsilon$.  It has been shown \cite{p84} that, if the density perturbations
have a Gaussian distribution, the {\it average} density profile around
a peak in the density distribution is given simply by
\begin{equation}
<\delta(r)> = \delta(0) \frac{\zeta(r)}{\zeta(0)}
\label{adpr}
\end{equation}
where $\delta(\vec{r}) \equiv \delta\rho(\vec{r})/\rho$ and
$\zeta(r) \equiv <\delta(\vec{r}) \delta(0)>$ is the 2-point correlation
function.  The latter is related to the power spectrum $P(k)$ by
\begin{equation} 
\zeta(r) =  \int \exp (i\vec{k}.\vec{r}) P(k) d^3 k.
\label{2pt}
\end{equation}
The power spectrum is the product $P(k) = A k^nT^2(k)$ where $A k^n$
is the primordial spectrum, taken for simplicity to be a power law, and 
$T(k)$ is the transfer function. For cold dark matter, the transfer 
function is given by \cite{bbks}
\begin{equation}
T(k)=\frac{\ln (1+2.34 q)}{2.34q} \times
\left[1+3.89q +(16.1q)^2+(5.46q)^3 +(6.71q)^4 \right]^{-1/4} \, ,
\label{tf}
\end{equation}
where  $q=k\,{\rm Mpc}/h^2$. The Harrison-Zel'dovich spectrum 
corresponds to $n=1$. Eqs.(\ref{adpr}) and (\ref{2pt}) imply 
that if $P(k) \sim k^{\alpha}$ on some momentum scale $k$, then 
$\zeta(r) \sim r^{-\alpha -3}$ and hence $\epsilon = (\alpha + 3)/3$
on the corresponding length scale $r=1/k$. We computed 
$\alpha = d\,ln P/d\,ln k$ for $n=1$ and plotted the resulting
$\epsilon(r)$ in Fig. \ref{fig:ps} for the relevant scales. The figure
suggests that $\epsilon$ is of order 0.2 - 0.3 on the galactic scale.

%%%%%%%%%%%%%%%%%%%%%%%%%%%%%%%%%%%%%%%%%%%%%%%%%%%%%%%%%%%%%%%%%%%%%%%%%%%%%

\section{Numerical Integration}
\label{sec:ni}

In this section, we present the results from numerically 
integrating Eqs. (\ref{mlb},\ref{mlc}) for various values
of the parameter $\epsilon$ and various angular momentum 
distributions, including no angular momentum, a single value of 
angular momentum and the distribution of Eq.(\ref{nj}).  The 
function $\lambda(\tau)$ gives us the phase-space distribution of
the particles through the equations:
\begin{mathletters}
\label{pss}
\begin{equation}
r(M_i,t)=r_*(M_i)\lambda \left({t\over t_*(M_i)}\right)
= R(t) \lambda(\tau) \tau^{-2/3-2/9\epsilon} \,\, ,
\label{pssa}
\end{equation}
\begin{equation}
v(M_i,t) = {d r(M_i,t) \over d t}=\frac{R(t)}{t}\,
\tau^{1/3-2/9\epsilon}{d\lambda \over d\tau} \, .
\label{pssb}
\end{equation}
\end{mathletters} 
If there is a distribution of angular momentum values, the 
functions $\lambda(\tau)$, $r(M_i,t)$ and $v(M_i,t)$ carry an 
index $k$ which we have suppressed here to avoid cluttering the 
equations.  

To solve Eq.(\ref{mlb}) for the particle trajectory 
$\lambda(\tau )$ we need to know the mass function ${\cal M}(\xi)$. 
This function, in turn, is given in terms of the trajectory 
$\lambda(\tau )$ by Eq.(\ref{mlc}) or, equivalently, by:
\begin{equation}
{\cal M}(\xi)=\sum_{n=1}\left(\tau_{2n-1}^{-2/3\epsilon}(\xi)-
\tau_{2n}^{-2/3\epsilon}(\xi) \right) \, ,
\label{explM}
\end{equation}
where the $\tau_j(\xi)$ correspond to the moments of time when the 
trajectory crosses radius $r=\xi R(t)$, i.e. they are the solutions
of $\lambda (\tau ) = \xi \tau^{2/3 + 2/9\epsilon}$. Following 
Fillmore and Goldreich \cite{fg84}, we solve Eqs.(\ref{mlb}) 
and (\ref{mlc}) simultaneously by a technique of successive iterations. 
Starting with some arbitrary mass profile ${\cal M}(\xi)$ (we took 
${\cal M}(\xi)=\xi^2$) we find $\lambda (\tau )$, which is then used 
to derive a new mass profile, from which a new trajectory is derived,
and so on.  The procedure is repeated till it converges.  We find that 
the mass profile changes very little after 5 iterations. Typically we 
run 10 iterations to get the final results. 

\begin{figure}
\psfig{file=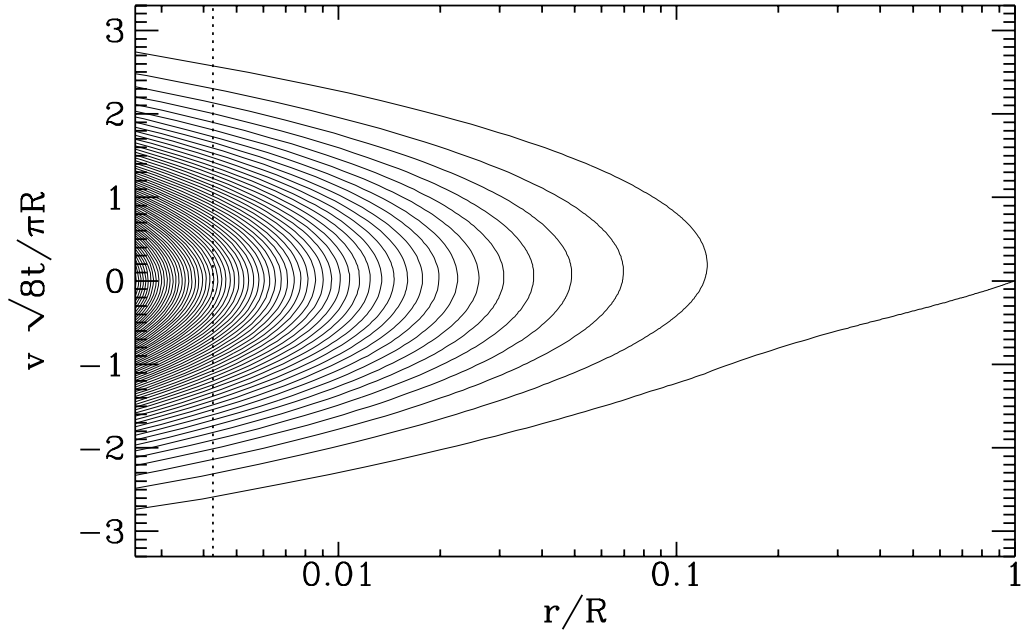,height=3.6in,width=5.6in}
\caption{The phase space distribution of halo dark matter particles at 
a fixed moment of time for the case $\epsilon =0.2$ and $j=0$. The solid 
lines represent occupied phase space cells.  The dotted line corresponds 
to the Sun's position if $h=0.7$.}
\label{fig:phs0}
\end{figure}

Fig. \ref{fig:phs0} shows the phase-space diagram for the case 
$\epsilon = 0.2$ and zero angular momentum.  The solid line in that figure 
shows the location of all the particles in phase space at a given time, 
i.e. it is the set of points $(r(M_i,t),v(M_i,t))$ for all $M_i$. 
The radial distances are normalized to the turnaround radius $R$ at time 
$t$ and the velocities are normalized to $\sqrt{G M(t) / R(t)} = 
\pi R(t) / {\sqrt{8} t}$ which is the rotation velocity at the 
turnaround radius.  Fig. \ref{fig:phs} shows the phase space diagram
for the case $\epsilon = 0.2$ and a single value of angular momentum 
$j= 0.2$. The particle trajectory $\lambda(\tau)$ for that case is 
shown in Fig. \ref{fig:ltau}.

\begin{figure}
\psfig{file=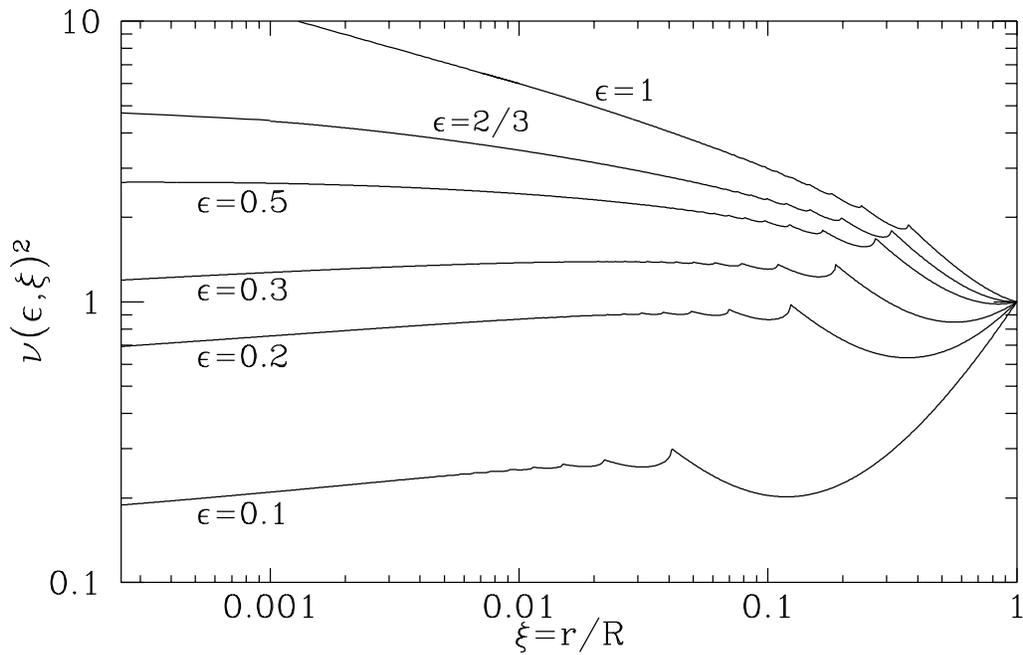,height=3.6in,width=5.6in}
\caption{Rotational velocity squared curves for different values of 
$\epsilon$ and $j=0$}
\label{fig:rcl0}
\end{figure}

A convenient way to show the mass distribution is by showing the rotation 
curve.  We define $\nu(\epsilon,\xi)$ by:
\begin{equation} 
v_{\rm rot}^2(r)= G M(r,t)/r\equiv \nu^2\left(\epsilon,{r\over R(t)}\right) 
{G M(t) \over R(t)}. 
\label{vrvr} 
\end{equation}
With this definition, we have $\nu(\epsilon,\xi=1)\, =1$.  The 
functions $\nu(\epsilon,\xi)^2$ obtained by numerical integration
are plotted in Fig. \ref{fig:rcl0} for various values of $\epsilon$ 
and $j=0$.

To fit the model to our galactic halo, we must choose values 
of the present turnaround radius $R \equiv R(t)$ and of $M \equiv M(t)$.
Equivalently, we may choose values of the present age $t$ and of $R$.  
$M$ is given in terms of $R$ and $t$ by Eq. (\ref{trta3b}).  $t$ is 
given in terms of the Hubble rate 
$H_0 = h\, 100$ km s$^{-1}$ Mpc$^{-1}$ by the relation $t^{-1}=3H_0/2$. 
We will use $h$ to state the age of the universe.  Then we fix $R$ in 
terms of $h$ by requiring that the model reproduce the measured value, 
$v_{\rm rot} =220$ km s$^{-1}$, of the rotation velocity in our galaxy.   
Let us call $\nu(\epsilon)$ the value of $\nu (\epsilon,\xi)$ in the 
flat part of the rotation curve, near $r=0.02R$ for $\epsilon < 0.4$; 
see Fig. \ref{fig:rcl0}. $\nu(\epsilon)$ is related to $v_{\rm rot}$ by 
Eq. (\ref{vrvr}). This implies: 
\begin{equation}
R\, h = 1.32\, \nu (\epsilon)^{-1}\, {\rm Mpc} \, .
\label{Rt}
\end{equation}
Table I gives $\nu (\epsilon)^2$, $R h$ and $M h$ for various values 
of $\epsilon$.   

Our distance to the galactic center is taken to be $8.5$ kpc and 
we define $\xi_s \equiv 8.5$ kpc$/R$. The contribution of the 
$n$-th velocity peak to the local halo density is given by:
\begin{equation}
\rho_n ={M \over 4\pi R^3 \xi_s^2}\, \, {6\tau_n^{2/3-4/9\epsilon} \over 
|9\epsilon\tau_n \dot{\lambda}_n-(6\epsilon +2)\lambda_n|} \, ,
\label{ifd}
\end{equation}
where $\dot{\lambda} \equiv d\lambda /d\tau$, and $\lambda_n$ and $\tau_n$ 
are the solutions of $\lambda (\tau ) = \xi_s \tau^{2/3 + 2/9\epsilon}$. 
The kinetic energy (in a frame of reference which is not rotating along 
with the galactic disk) per unit particle mass in the $n$-th peak is 
given by:
\begin{equation}
E_n =\frac{1}{2}v_n^2 = {R^2 \over 2 t^2}\, \tau_n^{2/3-4/9\epsilon} 
\left({j^2 \over \lambda_n^2} + \dot{\lambda_n}^2 \right) \, .
\label{ei}
\end{equation}
We shall express $E_n$ in units of $(300 \, {\rm km \, s}^{-1})^2 / 2$
when presenting our results.

We now discuss in greater detail the results specific to the different
types of angular momentum distributions used.

\subsection{Radial infall}
\label{ssec:ri}

Without angular momentum all particles pass through the origin, $r=0$,
at each oscillation. To avoid infinities in the numerical integration, a 
regulator at small $r$ is required.  The one which is most convenient
for us and which we use is to give a small amount of angular momentum  
to the dark matter particles.  We found $j^2= 10^{-6}$ to be small 
enough for our purposes.

Fig. \ref{fig:phs0} shows a typical phase space distribution.  
Fig. \ref{fig:rcl0} shows the rotation curves for various values 
of $\epsilon$.  An analytical treatment of the radial infall model 
using adiabatic invariants predicted  \cite{fg84} the behaviour of 
the density near the origin to be: 
$\rho \propto r^{-9\epsilon / (3\epsilon +1)}$ in the 
range $2/3 \leq \epsilon \leq 1$ and $\rho \propto r^{-2}$ in the 
range $0 < \epsilon \leq 2/3$.  These predictions agree very well 
with our results.  The rotation curves do indeed go to a constant near 
the origin when $0 < \epsilon \leq 2/3$ except for small logarithmic 
corrections.  The analytical treatment given in section V suggests 
the behaviour $\rho \sim 1/(r^2\sqrt{\ln (1/r)})$ for small $\epsilon$.  

\begin{figure}
\psfig{file=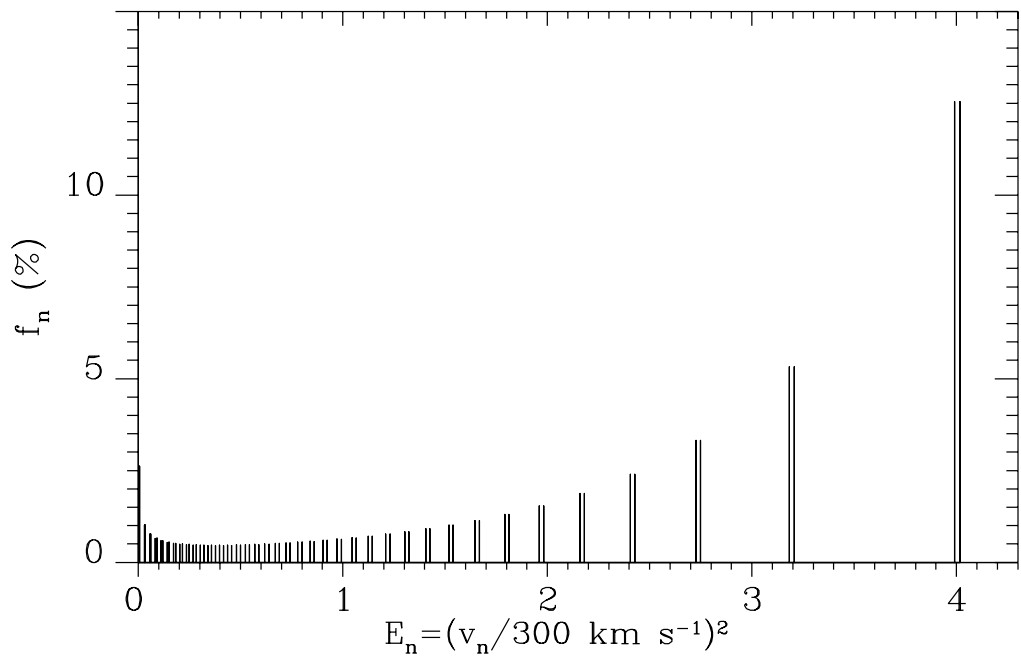,height=3.6in,width=5.6in}
\caption{The density fractions $f_n = \rho_n/\rho$ (in percent)
and the kinetic energies per unit particle mass $E_n$ of the peaks 
at the Sun's position, for $\epsilon=0.2$, $j=0$ and $h=0.7$. The peaks 
form pairs.  One member of each pair is due to particles with positive 
radial velocity and the other is due to particles with negative radial
velocity}
\label{fig:frs}
\end{figure}

Fig. \ref{fig:frs} shows the velocity peaks on Earth predicted by the 
radial infall model with $\epsilon =0.2$ and $h=0.7$. The rows 
labeled $\bar{j}=0.0$ in Table \ref{tbl3} give the density fractions 
and kinetic energies of the five most energetic incoming peaks
for the cases $\epsilon = 0.2$ and $1.0$, and $h = 0.7$. 
For each incoming peak there is an outgoing peak with approximately 
the same energy and density fraction.  We find that, in the radial 
infall model, the sizes of the two peaks due to particles falling in 
and out of the galaxy for the first time are large, each containing 
of order 10\% of the local halo density for $\epsilon$ in the standard 
CDM model inspired  range of 0.15 - 0.4. As was emphasized already, this 
spectrum is unrealistic because angular momentum has a non-negligible
effect upon the peak sizes.

\subsection{Single non-zero value of angular momentum.}
\label{sec:niri}

\begin{figure}
\psfig{file=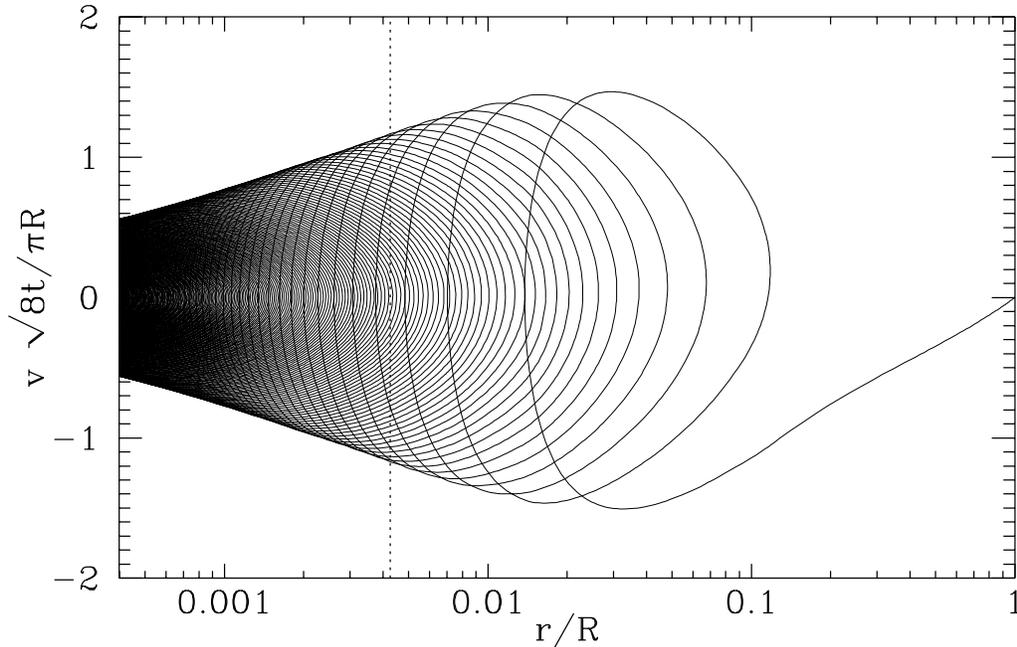,height=3.6in,width=5.6in}
\caption{The phase space distribution of the dark matter particles in
the case $\epsilon=0.2, h=0.7$ and a single value of angular momentum
$j=0.2$}
\label{fig:phs}
\end{figure}

\begin{figure}
\psfig{file=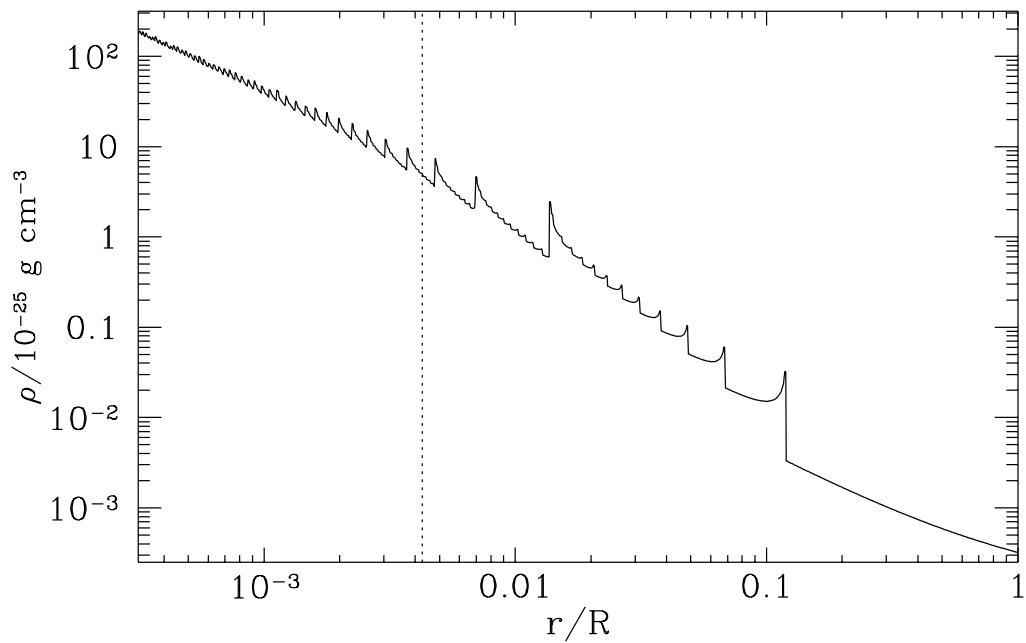,height=3.6in,width=5.6in}
\caption{Density profile for the case $\epsilon =0.2, h=0.7$ and a single
value of angular momentum $j=0.2$.}
\label{fig:rho}
\end{figure}

Fig. \ref{fig:phs} shows the phase space diagram for the case 
$\epsilon = 0.2$ and $j = 0.2$.  In the model with a single 
non-zero value of angular momentum, the halo distribution has a set 
of inner caustics in addition to the usual outer caustics.  However, 
the inner caustics are spheres in the model whereas they are rings 
for a physical halo.  The density profile for the model is shown in 
Fig. \ref{fig:rho}.  There is a break in the logarithmic slope near 
the first inner caustic, near $r = 0.01 R$ in the figure.  
For $\epsilon < 2/3$ we find $\rho \propto r^{-2}$ 
outside the first inner caustic (the same as with $j = 0$)
but $\rho \propto r^{-\gamma}$ inside with $\gamma = 9 \epsilon
/(3 \epsilon + 1)$.  This observation is related to the fact that 
we find the function $\lambda (\tau)$ to be oscillating 
with constant amplitude and constant period for large $\tau$ when
$j \neq 0$; see Fig. \ref{fig:ltau}.  This can happen only if 
$\tau^{2/3\epsilon}{\cal M}\left(\lambda/\tau^{2/3+2/9\epsilon}\right)$ 
in Eq. (\ref{mlb}) is independent of $\tau$ at large $\tau$. For 
${\cal M}(r) \propto r^\alpha$, this implies 
$\alpha =  3/(3\epsilon +1)$ or 
$\gamma = 3 -\alpha = 9\epsilon/(3\epsilon +1)$.  We find that 
the exponent $\alpha$ does not depend upon $j$, but the radius 
$r_c$ at which the break in power law behaviour occurs  does depend 
upon $j$: the smaller $j$, the smaller $r_c$. 

These observations may be understood as follows.  When angular momentum 
is present, the contribution of a single phase-space sheet to the density 
profile $\rho(r)$ is non-singular near $r=0$.  In that case, as a result 
of self-similarity the mass profile $M(r)$, at $r$ small enough
that the influence of the first one or two phase-space sheets may 
be neglected, has the same functional dependence upon $r$ as the 
dependence of $M$ upon $R$ after eliminating $t$ from the expressions 
for $M(t)$ and $R(t)$ in Eqs. (\ref{trta3}).  This yields 
$M \propto r^{3/(3\epsilon+1)}$.  The fact that $\rho(r)$ behaves as 
a negative power of $r$ near $r=0$ is in agreement with the results 
of N-body simulations \cite{dcwz}.  The $\epsilon$-dependence of 
this power law was already known to White and Zaritsky \cite{sw}.  

The spectrum of velocity peaks that the model with a single value
of angular momentum predicts is unrealistic.  In particular, there 
are in this model no peaks on Earth associated with particles falling 
in or out of the galaxy for the $n$-th time with $n$ small, because such 
particles have too much angular momentum to reach the Earth radius. In 
the case of Fig. \ref{fig:phs}, there are only peaks for $n > 3$.  But as 
we argued at length in section II, there are in reality peaks on Earth 
due to particles falling in and out of the galaxy for the $n$-th time 
with $n = 1,2,3, ...$  Angular momentum reduces the sizes of the peaks 
with small $n$ but does not suppress them completely. 

\subsection{Distribution of angular momenta.}

This model describes a physical halo averaged over all possible 
orientations.  The particles are assumed to have the distribution 
of angular momenta given in Eq.(\ref{nj}).  For the purpose of 
numerical integration, the angular momentum was discretized with a  
spectrum of four hundred values.  The phase space diagram 
is like the one of Fig. \ref{fig:phs} except that there are 
four hundred such diagrams superimposed on one another, one 
for each value of $j$.

\begin{figure}
\psfig{file=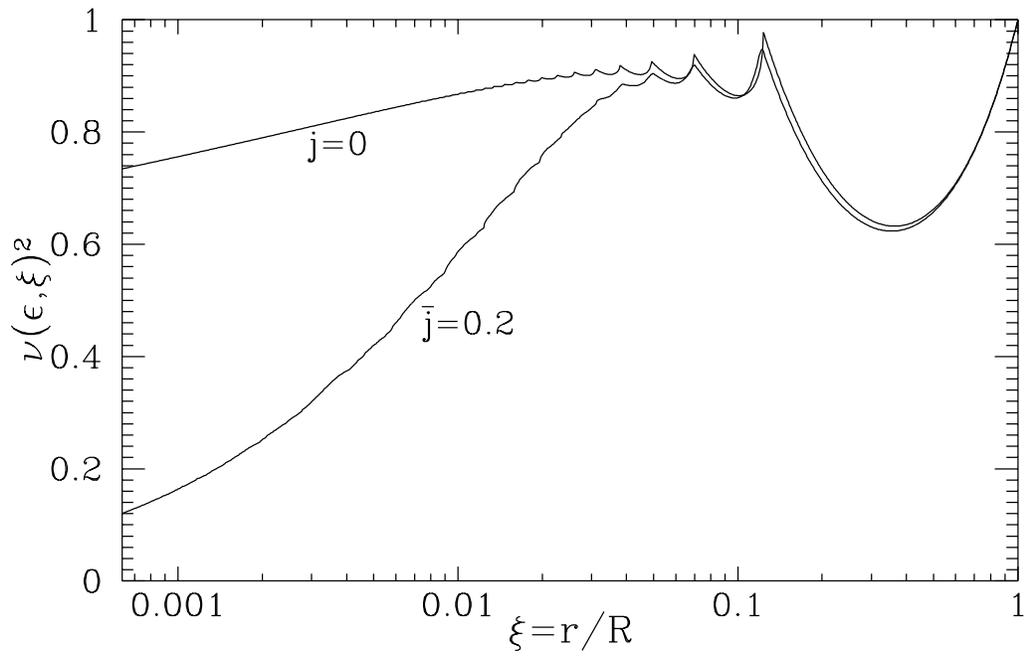,height=3.6in,width=5.6in}
\caption{Rotational curves for the case $\epsilon =0.2$, with and
without angular momentum.}
\label{fig:rvj02}
\end{figure}

As was mentioned earlier, angular momentum has the effect of 
making the contribution of the halo to the rotational velocity 
go to zero at the galactic
center as shown by Fig. \ref{fig:rvj02}.  We define the 
``effective core radius'' $b$ to be such that $\nu^2 (\epsilon,b/R) 
= \nu^2 (\epsilon)/2$. We find that near $r=0$, the density 
$\rho(r) \propto r^{-\gamma}$ with $\gamma = 9 \epsilon /(3 \epsilon +1)$
as in the case of a single value of angular momentum.   This behaviour
is expected for the same reason as we gave for that case.  The only 
change is that the transition between the region at large $r$ where 
$\rho(r)\propto r^{-2}$ and the region at small $r$ where  
$\rho(r) \propto r^{-\gamma}$ is smoother
now because the critical radius $r_c$ where the transition occurs 
depends upon $j$ and there is now a distribution of $j$ values.
  
\begin{figure}
\psfig{file=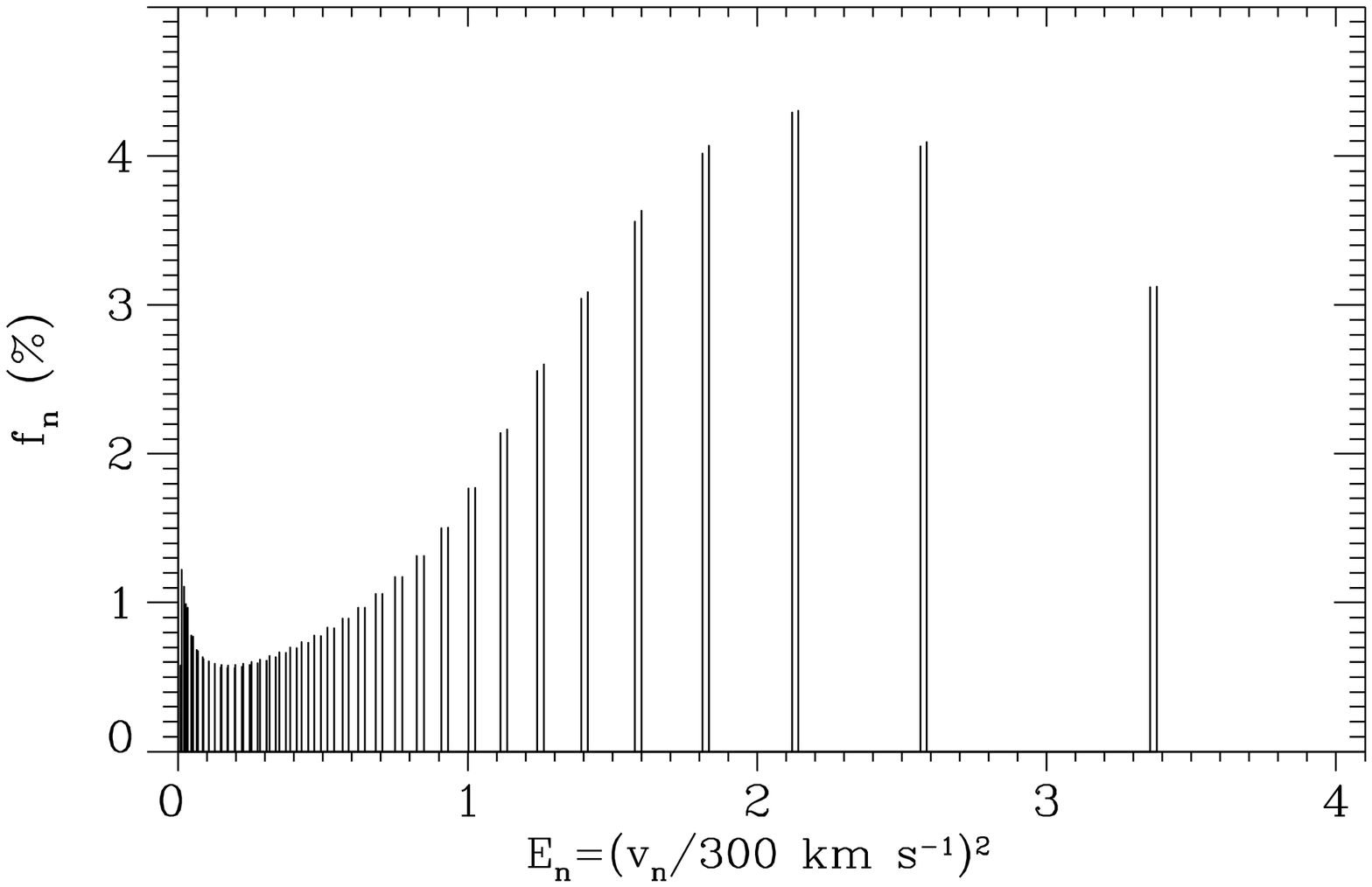,height=3.6in,width=5.6in}
\caption{The same as Fig.~\ref{fig:frs}  but $\epsilon=0.2, h=0.7$
and $\bar{j}=0.2$.}
\label{fig:vpd02}
\end{figure}

\begin{figure}
\psfig{file=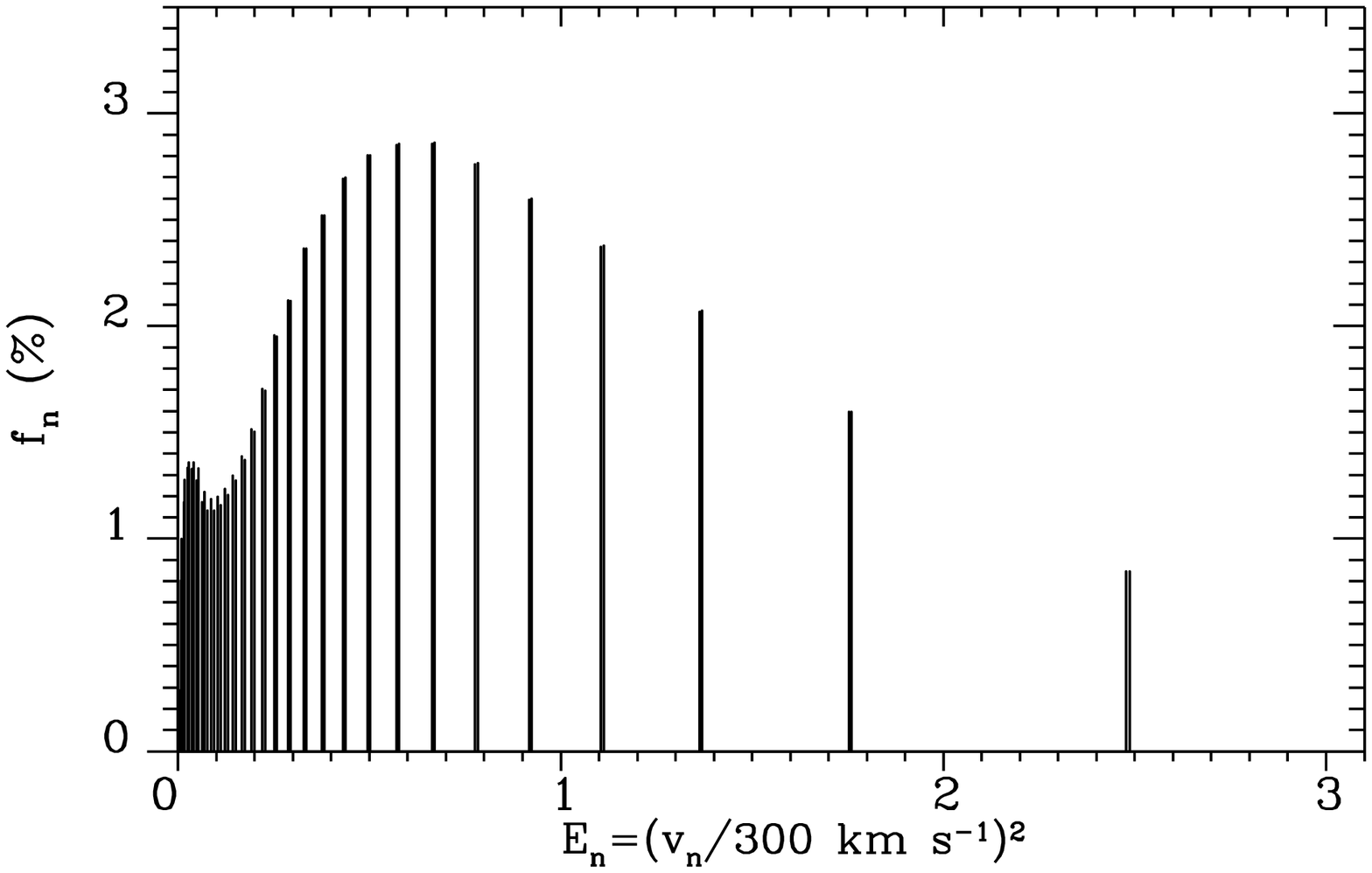,height=3.6in,width=5.6in}
\caption{The same as Fig.~\ref{fig:frs}  but $\epsilon=0.2, h=0.7$
and $\bar{j}=0.4$.}
\label{fig:vpde02j04}
\end{figure}

\begin{figure}
\psfig{file=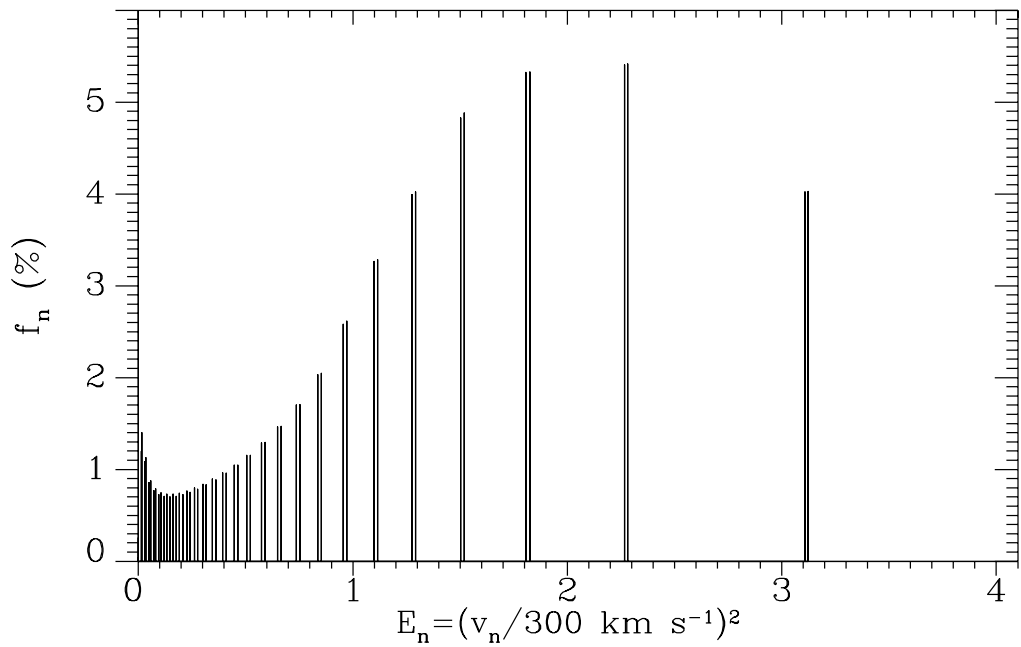,height=3.6in,width=5.6in}
\caption{The same as Fig.~\ref{fig:frs}  but for $\epsilon=0.15, h=0.7$
and $\bar{j}=0.2$.}
\label{fig:vpd015}
\end{figure}

\begin{figure}
\psfig{file=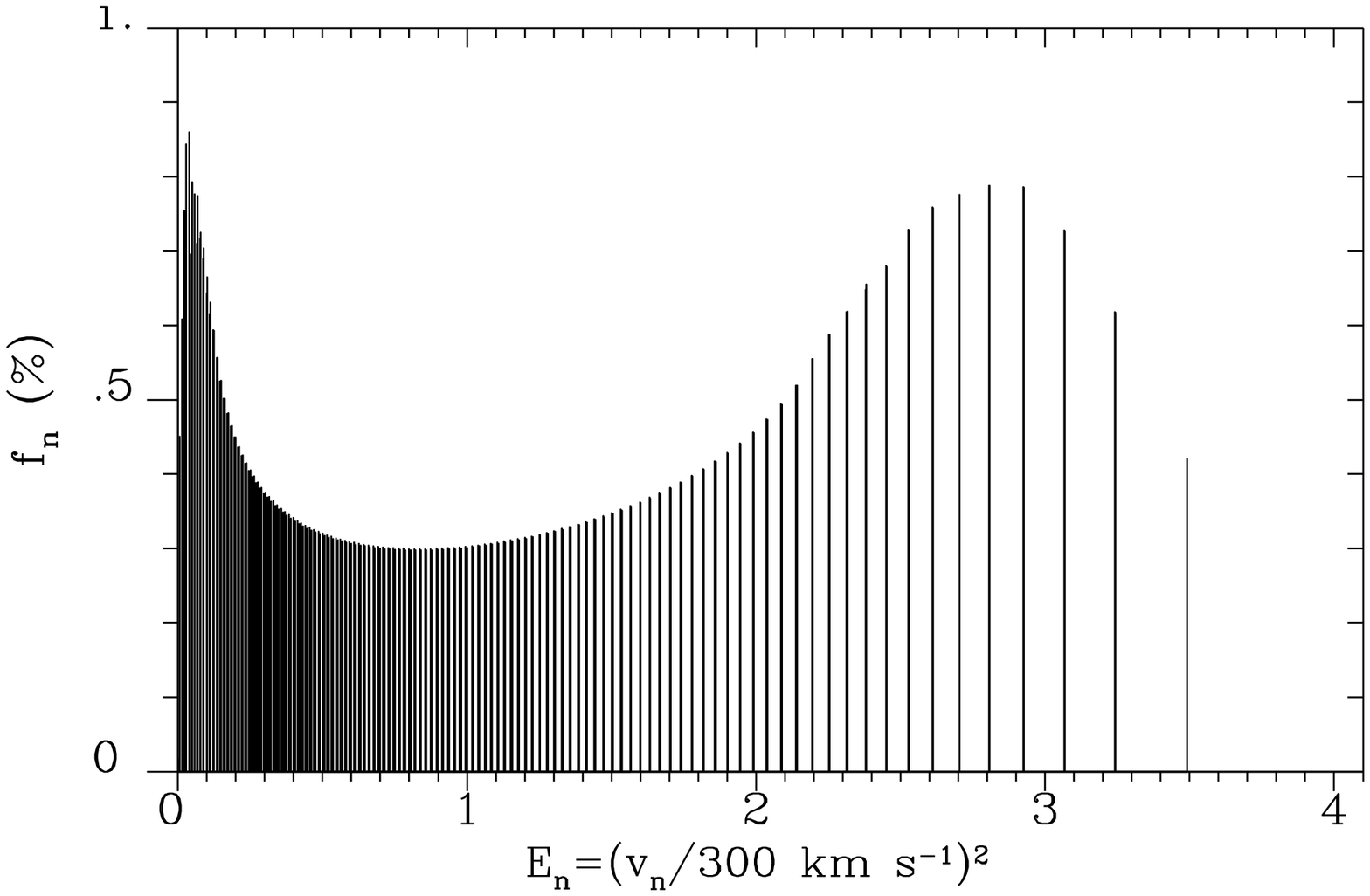,height=3.6in,width=5.6in}
\caption{The same as Fig.~\ref{fig:frs}  but $\epsilon=1, h=0.7$ and
$\bar{j}=0.2$.}
\label{fig:vpd1}
\end{figure}

The velocity peaks for $\bar{j}=0.2$, $\epsilon =0.2$ and $h=0.7$ 
are shown in Fig. \ref{fig:vpd02}.  The velocity peaks for 
the same choice of parameters except $\bar{j}=0.4$ are shown 
in Fig. \ref{fig:vpde02j04} to indicate the sensitivity of the 
peaks upon $\bar{j}$. The spectrum of velocity peaks is also sensitive 
to the value of $\epsilon$.  It is shown for the cases $\epsilon =0.15$ 
and $\epsilon =1$ on Figs. \ref{fig:vpd015} and \ref{fig:vpd1} respectively.

Table \ref{tbl3} gives the values of the current turn-around radius 
$R$, the effective core radius $b$, the halo density at our location 
$\rho(r_s)$, and the density fractions and kinetic energies 
of the five most energetic incoming peaks for various values of 
$\epsilon$, $\bar{j}$ and $h$. 

\subsection{Self-similar infall with baryons}
\label{sec:bi2}

\begin{figure}
\psfig{file=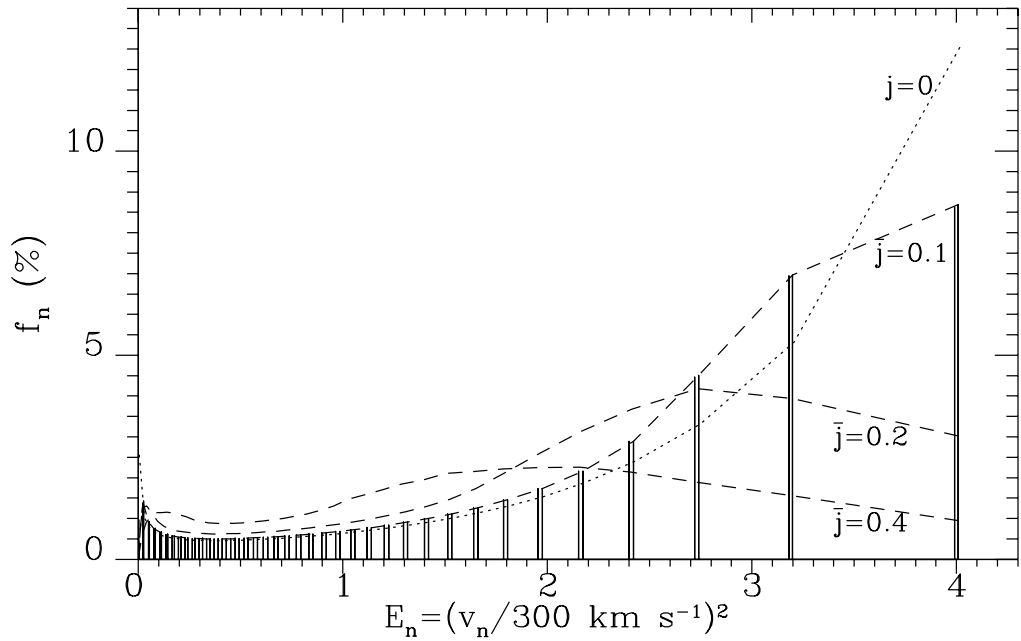,height=3.6in,width=5.6in}
\caption{The same as Fig.~\ref{fig:frs} but including the contribution 
of baryons to the galactic gravitational potential. $\epsilon=0.2$
and $h=0.7$ in all cases.  The peaks are shown explicitly for $\bar{j}=0.1$. 
The dashed lines go through the tops of the peaks for the cases $\bar{j}=0.2$ 
and $\bar{j}=0.4$ while the dotted line corresponds to $j=0$.}
\label{fig:vpbi}
\end{figure}

\begin{figure}
\psfig{file=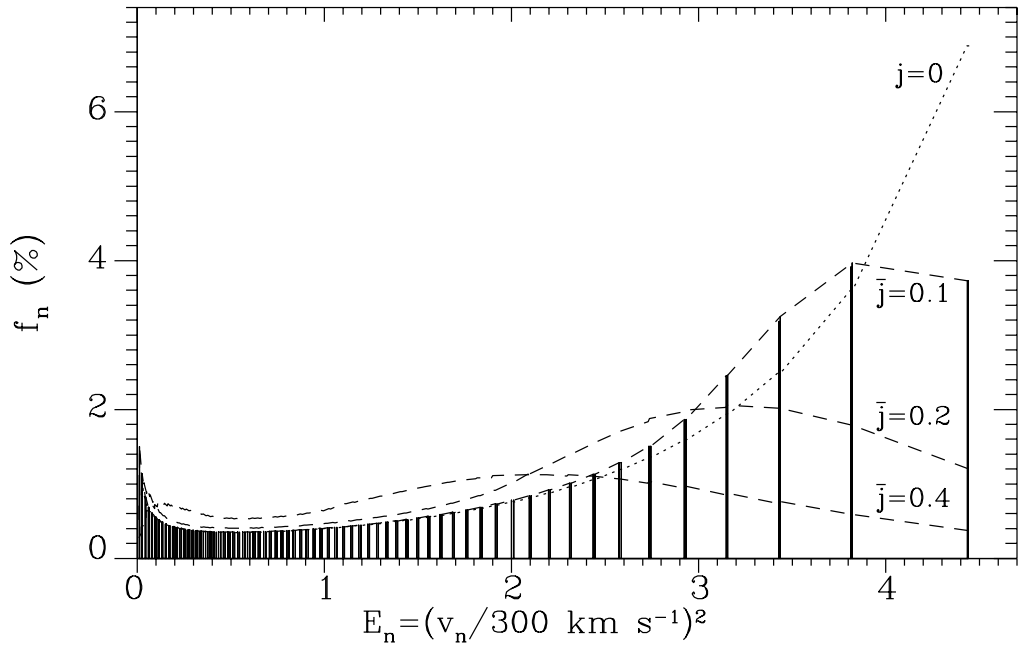,height=3.6in,width=5.6in}
\caption{The same as Fig.~\ref{fig:vpbi} but for $\epsilon=0.4$.}
\label{fig:vpbi_e04}
\end{figure}

As was discussed earlier, the dark matter phase space distribution
for the input parameters $\epsilon$, $h$ and $\bar{j}$ is obtained in this 
case by solving Eq. (\ref{mlc}) using the mass distribution ${\cal M}(\xi)$
for the same values of $\epsilon$ and $h$ but $j=0$.  The resulting velocity 
peaks are shown in Fig. \ref{fig:vpbi} for $\epsilon =0.2$, $h=0.7$ and 
different values of $\bar{j}$.  Since the particles for different values of
$\bar{j}$ but the same values of $\epsilon$ and $h$ are all moving in the 
same gravitational potential, the kinetic energies per unit particle mass
$E_n$ of the peaks depend only very weakly upon $\bar{j}$.  So, we have 
combined the spectra for different $\bar{j}$ values in one figure.  Note 
that the $E_n$ are larger in this case than in the case of 
Figs. \ref{fig:vpd02} and \ref{fig:vpde02j04} because the gravitational 
well near the center of the galaxy is deeper.  Fig. \ref{fig:vpbi_e04} 
shows the results for $\epsilon=0.4$, $h=0.7$.

\section{An analytical treatment.}
\label{sec:a1p}

In this section, we derive analytical expressions for the sizes and
locations of the velocity peaks. The treatment involves the following 
approximations:
\begin{enumerate}
\item the mass distribution, including the contributions from both
baryons and dark matter, is taken to be ${\cal M}(\xi)=\xi$ for 
$ 0 \le \xi \le 1$
\item the dimensionless angular momentum $j$ values are assumed to be small
\item our distance $r_s$ to the galactic center is assumed to be small 
compared to the oscillation amplitudes of the particles in the peaks 
under consideration. 
\end{enumerate}
The method of adiabatic invariants will be used to obtain the motion of 
the dark matter particles. 

We first treat the case of a single angular momentum value, i.e the model 
described in subsection III.B.1. Since ${\cal M}(\xi )=\xi$, the equation 
of motion in rescaled variables of a particle with angular momentum $j$ is: 
\begin{equation} 
{d^2\lambda \over d\tau^2}= {j^2 \over \lambda^3}-{\pi^2 \over
8\lambda}\tau^{\frac{2}{3}(\frac{2}{3\epsilon}-1)}  \,\, .
\label{s5d2ll2}
\end{equation}
Let $\lambda_M$  and $\lambda_m$  be respectively the amplitude of 
oscillation and the distance of closest approach at time $\tau$.
In the spirit of the method of adiabatic invariants, $\lambda_M$  
and $\lambda_m$ are assumed to be slowly varying functions of $\tau$. 
This is a valid assumption for all oscillations except the first one.  
The first oscillation is only marginally adiabatic.  For given  
$\lambda_M$, the velocity at time $\tau$ and position $\lambda$ is:
\begin{equation}
{d\lambda \over d\tau}=\pm \sqrt{{\pi^2 \over 4}
\tau^{\frac{2}{3}(\frac{2}{3\epsilon}-1)} \ln \left( \frac{\lambda_M}{\lambda}
\right) - j^2 \left( \frac{1}{\lambda^2}-\frac{1}{\lambda_M^2}\right)}
\,  \, .
\label{s5e2}
\end{equation}
The adiabatic invariant is:
\begin{eqnarray}
&&I(\tau,\lambda_M)=\int_{\lambda_m}^{\lambda_M} d\lambda |d\lambda/d\tau| 
\nonumber \\
&&= \int_{\lambda_m/\lambda_M}^1 dx \sqrt{-{\pi^2 \over 4}
\tau^{\frac{2}{3}(\frac{2}{3\epsilon}-1)} \lambda_M^2 \ln x 
 - j^2 \left( \frac{1}{x^2}- 1 \right)} \, .
\label{s5e3}
\end{eqnarray}
In the limit of small $j$, $\frac{\lambda_m}{\lambda_M}$  may be neglected 
and we obtain
$\lambda_M(\tau)^2  \tau^{\frac{2}{3}(\frac{2}{3\epsilon}-1)}= $ constant. 
Since  $\lambda_M(1)=1$, we have:
\begin{equation}
\lambda_M(\tau) = \tau^{-\frac{1}{3}(\frac{2}{3\epsilon}-1)} \, .
\label{s5e4}
\end{equation}
The period of oscillation is:
\begin{eqnarray}
&&T(\tau,\lambda_M)
= 2\int_{\lambda_m}^{\lambda_M} d\lambda |d\lambda/d\tau|^{-1} 
\nonumber \\
&&= 2 \lambda_M^2 \int_{\lambda_m/\lambda_M}^1 dx \left[-{\pi^2 \over 4}
\tau^{\frac{2}{3}(\frac{2}{3\epsilon}-1)} \lambda_M^2 \ln x 
 - j^2 \left( \frac{1}{x^2}- 1 \right)\right]^{-1/2}  \, .
\label{s5e5}
\end{eqnarray}
Using Eq. (\ref{s5e4}), we have in the limit of small $j$:
\begin{equation}
T(\tau)=\frac{4\lambda_M}{\pi\tau^{\frac{1}{3}(\frac{2}{3\epsilon}-1)}} 
\int_0^1 {dx \over \sqrt{- \ln x}}={4 \over \sqrt{\pi}} 
\tau^{-\frac{2}{3}(\frac{2}{3\epsilon}-1)} \, .
\label{s5e6}
\end{equation}
Let us introduce the ``phase'' $\varphi (\tau)$ :
\begin{equation}
d \varphi =\pi {d\tau \over T(\tau)} \, \, .
\label{s5e7}
\end{equation}
Using Eq. (\ref{s5e6}) and setting $\varphi(1)=0$, we have:
\begin{equation}
\varphi(\tau )= 9 \epsilon\, {\pi \sqrt{\pi} \over 4} \,
{\tau^{\frac{1}{3}(\frac{4}{3\epsilon}+1)}-1 \over 4+3\epsilon} \, .
\label{s5e8}
\end{equation}
The times $\tau_n$  at which the particle passes by the solar 
radius $r_s$  are given by:                                                    
\begin{eqnarray}
\varphi(\tau_n) &=& \frac{\pi}{2} (2n-1) + o(r_s/R)
\label{s5e9}
\end{eqnarray}
There are two peaks for given $n$, one ingoing and the 
other outgoing.  The differences between the properties of the two 
peaks are of order $o(r_s/R)$ and are neglected.
                                                                               
The contribution of each of the two $n$-th velocity peaks to the 
local density is given by:
\begin{equation}
\rho_n = {M \over 4\pi R r_s^2}\,{6 \tau_n^{\frac{2}{3}(1-\frac{2}
{3\epsilon})}\over |9 \epsilon \tau_n \frac{d\lambda}{d\tau}(\tau_n) 
-(6\epsilon+2)\lambda (\tau_n) |} \,  ,
\label{s5e10}
\end{equation}
where
\begin{equation}
\lambda(\tau_n)=\frac{r_s}{R}\, \tau_n^{\frac{2}{3}+\frac{2}{9\epsilon}}
\label{s5e11}
\end{equation}
and $ \frac{d\lambda}{d\tau}(\tau_n) $ is given by Eqs. (\ref{s5e2}) 
and (\ref{s5e4}). We neglect the second term in the denominator of Eq. 
(\ref{s5e10}) since it is $o(r_s/R)$ relative to the first term. 
Combining everything, we have the following estimate for the peak sizes 
due to dark matter particles with a single value of angular momentum $j$:
\begin{equation}
\rho_n = {M \over 6\pi R r_s^2\epsilon}\,
\left[ \frac{\pi^2}{4}\tau_n^{4/3\epsilon} \ln \left(\frac{R}
{r_s \tau_n^{\frac{1}{3}+\frac{4}{9\epsilon}}}\right)-j^2\frac{R^2}
{r_s^2}\tau_n^{\frac{4}{9\epsilon}-\frac{2}{3}}\right]^{-1/2}    \,  ,
\label{s5e12}
\end{equation}
where
\begin{equation}
\tau_n=\left[\frac{2}{\sqrt{\pi}}(2n-1)\left(\frac{4}{9\epsilon}+\frac{1}
{3} \right)+1\right]^{\frac{9\epsilon}{4+3\epsilon}} \,  .
\label{s5e13}
\end{equation}
Note that $\rho_n$ is the local energy density contributed by {\em each} 
of the two $n$-th peaks.  When the expression under the square root in 
Eq. (\ref{s5e12}) is negative, one must set the corresponding $\rho_n=0$ 
since this corresponds to the situation where the particles have too 
much angular momentum, and hence too large a distance of closest approach 
to the galactic center, to reach the solar radius. The kinetic energy per 
unit particle mass of the dark matter particles in the $n$-th peaks is:
\begin{eqnarray}
E_n &=& \frac{1}{2}\left[ \left(\frac{d\lambda}{d\tau}\right)^2_n +
\frac{j^2}{\lambda^2_n}\right] \left(\frac{r_*}{t_*}\right)^2_n
\nonumber \\
&=& \frac{\pi^2}{8}\left(\frac{R}{t}\right)^2 \ln \left( \frac{R}
{r_s\tau_n^{\frac{1}{3}+\frac{4}{9\epsilon}}}\right) + o(r_s/R)^2  \, .
\label{s5e14} 
\end{eqnarray}
$R$ and $t$ are determined in terms of $h$ by $t=2/3H_0$ and Eq. (\ref{Rt})
as before. 

For the case $\epsilon=0.2$, $j=0.2$, $h=0.7$, the quantity under 
the square root in Eq. (\ref{s5e12}) is negatve and hence $\rho_n=0$
for $n$=1,2 and 3.  This is consistent with the phase space 
distribution shown in Fig. \ref{fig:phs} which was obtained by 
numerical integration.  We found Eqs. (\ref{s5e12} - \ref{s5e14}) 
to be  consistent with the results from numerical integration at 
the 30\% level or so.  The agreement is worse for the case of zero 
angular momentum, probably because the logarithmic singularity 
of the potential at the origin in that case makes the motion 
non-adiabatic there.  Also, when comparing results for the $E_n$ values, 
one should allow for an overall shift between the two calculations 
due to a change in the depth of the gravitational potential at the 
solar radius.  Indeed, the analytical calculation has the mass 
distribution ${\cal M}(\xi ) =\xi$ which implies a perfectly flat 
rotation curve, whereas the rotation curves for the numerical calculations
are as shown in Fig.\ref{fig:rcl0}. 

\begin{figure}
\psfig{file=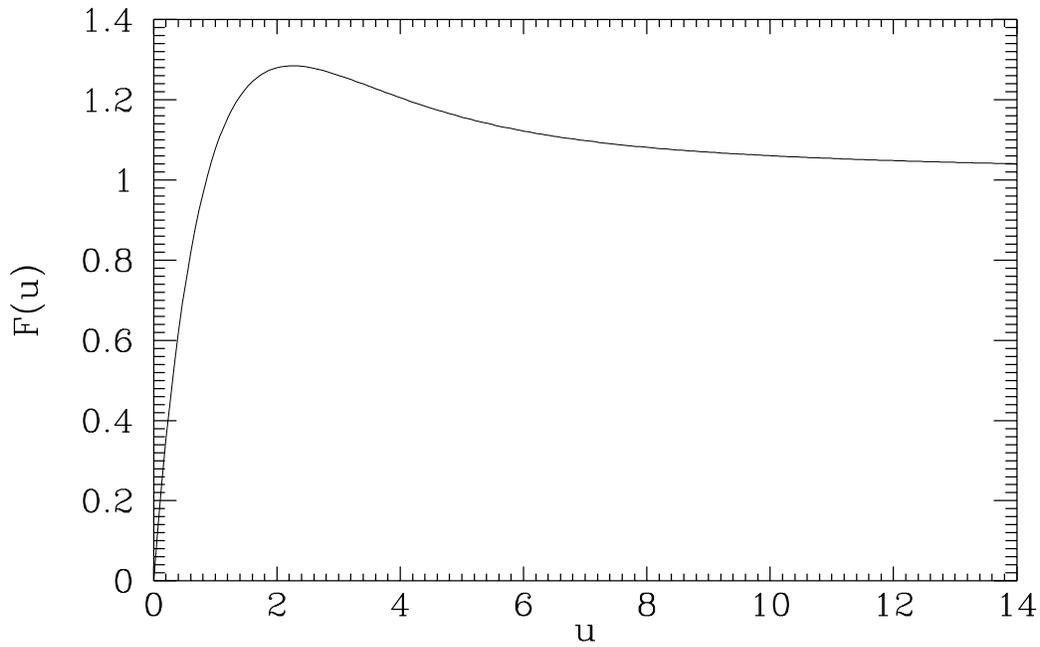,height=3.6in,width=5.6in}
\caption{The function $F(u)$ defined in Eq. \ref{s5e17}.}
\label{fig:fu}
\end{figure}

The formalism readily accomodates a distribution $dn/dj$  of angular 
momenta.  Eq. (\ref{s5e14}) for the peak kinetic energies still applies. 
The expected peak sizes, in the sense of an average over all 
locations at distance $r_s$ from the galactic center, are given 
by the convolution of $dn/dj$ with the expression, Eq. (\ref{s5e12}), 
for the peak sizes when there is a single value of angular momentum. Thus:
\begin{equation}
\rho_n = {M \over 6\pi R r_s^2\epsilon}\, \int_0^{j_n} dj \frac{dn}{dj}
\left[ \frac{\pi^2}{4}\tau_n^{4/3\epsilon} \ln \left(\frac{R}
{r_s \tau_n^{\frac{1}{3}+\frac{4}{9\epsilon}}}\right)-j^2\frac{R^2}
{r_s^2}\tau_n^{\frac{4}{9\epsilon}-\frac{2}{3}}\right]^{-1/2}\,  ,
\label{s5e15}
\end{equation}
where $\tau_n$ is given by Eq. (\ref{s5e13}) as before and $j_n$ is 
the maximum value of $j$ for which the argument of the square root 
is positive:
\begin{equation}
j_n= {\pi \over 2} \tau_n^{\frac{4}{9\epsilon}+\frac{1}{3}} \,
{r_s \over R} \sqrt{\ln \left(\frac{R}
{r_s \tau_n^{\frac{1}{3}+\frac{4}{9\epsilon}}}\right)} \, .
\label{jn}
\end{equation}
For the angular momentum distribution of Eq. (3.26), one has 
\begin{equation}
\rho_n={M \over 6\pi R^2 r_s \epsilon}\, 
{\tau_n^{\frac{1}{3}-\frac{2}{9\epsilon}} \over j_n}\, 
F\left((j_n/j_0)^2\right) \, ,
\label{s5e16}
\end{equation}
where
\begin{equation}
F(u) \equiv u \int_0^1 dx e^{-ux}(1-x)^{-1/2} \, .
\label{s5e17}
\end{equation}
A graph of $F(u)$ is shown in Fig. \ref{fig:fu}. Table 3 shows the peak 
sizes predicted by Eq.(\ref{s5e16}) for the first eight peaks in 
the case $\epsilon=0.2$, $\bar{j}=0.2$ and $h=0.7$. The peak sizes agree 
with the results of the numerical integration, given in the fifth line of 
Table 3, to within 15\%. The peak energies also agree within 15\% after 
one has subtracted an overall shift caused by the fact that the 
gravitational potential at our location is considerably deeper in the 
analytical calculation than in the numerical one.

\section{Conclusions}
\label{sec:co}

Motivated by the prospect that the spectrum of dark matter particles
on Earth may some day be measured in a direct detection experiment, 
we have endeavoured to obtain predictions for the properties of that 
spectrum.  Previously, it had generally been assumed that the spectrum 
is isothermal.  In contrast, we find that there will be large 
deviations from an isothermal spectrum in the form of peaks in velocity 
space associated with particles that are falling in and out of the Galaxy 
for the first time and with particles that have fallen in and out of 
the Galaxy only a small number of times in the past.

To obtain estimates for the velocity magnitudes of the particles in the
peaks and for the contributions of the individual peaks to the dark matter
local density, we have used the secondary infall model of galactic halo
formation.  We have generalized the extant version of that model to 
include the effect of angular momentum.  We forced spherical symmetry 
onto the model with angular momentum by averaging over all possible 
orientations of a physical halo.  As a result, the model can only make 
predictions for the average properties of the velocity peaks, the average 
being over all locations at the same distance from the galactic center 
as we are.  We found that the model with angular momentum has self-similar 
solutions if the angular momentum distribution, as well as the initial 
overdensity profile, has a particular scale-free form.  The self-similar 
solutions were analyzed in detail numerically and analytically.

The model produces a good overall fit to what is known about galaxies 
like our own.  It produces flat rotation curves when the parameter 
$\epsilon$ is in the range 0 to 2/3.  The expected value of that 
parameter in models of large scale structure formation with cold 
dark matter and a flat (Harrison- Zel'dovich) spectrum of primordial 
density perturbations is $\epsilon \sim 0.25$.  The model implies 
a relationship (cfr. Eq. (\ref{Rt}) and Table I) between the 
current turn-around radius $R$, the present age of the universe,
the galactic rotation velocity and the paramemter $\epsilon$.  This
relationship is consistent with observations and $\epsilon \sim 0.25$.  

In the model the galactic rotation curve is approximately 
flat all the way out to the turn-around radius $R$. $R$ is of order 1-2 Mpc 
for our galaxy.  So far, the rotation curves of individual galaxies 
have been measured, and have been found to be flat, up to distances of 
order 100 kpc, implying masses of order $10^{12} M_\odot$ or larger.  The 
discovery of flat rotation curves \cite{flat} caused a well-known 
revolution in our concept of galaxies.  Prior to these measurements 
galaxies were thought to have size of order 10 kpc and mass of order 
$10^{11} M_\odot$.  The model implies that galaxies like our own are 
of order 20 times bigger even than the minimum size implied by the 
rotation curve measurements.  The size is 1.8 Mpc and the mass is 
$1.7 \times 10^{13} M_\odot$ for the case $\epsilon=0.25, h=0.7$.  
See Table I for other cases.  With galaxies that massive, one has 
$\Omega \simeq 1$ in galaxies. 

We found that the effect of angular momentum is to deplete the inner 
parts of the halo with the result that the halo contribution to the 
rotation curve goes to zero at $r=0$. The model establishes a direct 
relation between the amount of angular momentum and the effective core 
radius $b$, defined as the radius at which the halo contributes half 
of the rotation velocity squared. Table \ref{tbl3} gives the turn-around
radius $R$, the effective core radius $b$ and the local halo density
$\rho$ as a function of the input parameters: $\epsilon$, the average
amount of dimensionless angular momentum $\bar{j}$ and the Hubble 
parameter $h$.  The observed value (220 km/s) of our galaxy's rotation 
velocity and our distance to the galactic center (8.5 kpc) were used 
as fixed input parameters. 

Table \ref{tbl3} also gives the average values of the peak sizes as fractions
of the local halo density $\rho$ and the kinetic energies per unit particle 
mass of the particles for the first five incoming peaks as a function of 
the variable input parameters $\epsilon, \bar{j}$ and $h$.  For each 
incoming peak there is an outgoing peak with approximately the same kinetic
energy and average local density. Let us emphasize again that the peak 
sizes given are averages over all locations at the same distance (8.5 kpc)
from the galactic center as we are. It is not possible to make more
precise predictions for the peak sizes on Earth without assuming a 
particular angular momentum field for the infalling dark matter.

\acknowledgments
We thank J. Ipser, S. Tremaine and D. Eardly for useful discussions.
This research was supported in part by DOE grant DE-FG05-86ER40272
at the University of Florida, DOE grant DE-AC02-76ER01545 at Ohio State 
and the DOE and NASA grant NAGW-2381 at Fermilab.

%%%%%%%%%%%%%%%%%%%%%%%%%%%%%%%%%%%%%%%%%%%%%%%
\def\baselinestretch{1.}

\begin{table}
\caption{$\nu^2(\epsilon)$, $Rh$ (in units of Mpc) and $Mh$ 
(in units of $M_\odot$) for different values of $\epsilon$.}
\begin{tabular}{cccc}
 $\epsilon$ &  $\nu^2(\epsilon)$ & $Rh$ & $Mh$  \\
\tableline
 0.1   & 0.25  & 2.6 & $1.1 \times 10^{14}$  \\
 0.15  & 0.6   & 1.7 & $3.2 \times 10^{13}$  \\
 0.2   & 0.9   & 1.4  & $1.8 \times 10^{13}$ \\
 0.25  & 1.15  & 1.23 & $1.2 \times 10^{13}$ \\
 0.3   & 1.35  & 1.14 & $9.5 \times 10^{12}$ \\
 0.35  & 1.5   & 1.08 & $8.1 \times 10^{12}$ \\
 0.4   & 1.7   & 1.02 & $6.8 \times 10^{12}$ \\
 0.45  & 1.8   & 0.98 & $6.1 \times 10^{12}$ \\
\end{tabular}
\label{tbl1}
\end{table}

\widetext

\begin{table}
\caption{Density fractions $f_n$ and kinetic energies per unit particle 
mass $E_n$ of the first five incoming peaks for various values of $\epsilon$,
$h$ and the average dimensionless angular momentum $\bar{j}$.  Also shown 
are the current turn-around radius $R$ in units of Mpc, the effective core 
radius $b$ in kpc, and the local density $\rho$ in units of
$10^{-25}$ g cm$^{-3}$. The $f_n$ are in percent and the
$E_n$ are in units of $0.5 \times (300 $ km s$^{-1})^2$. }
\begin{tabular}{ccccccccccc}
$\epsilon$ &  $\bar{j}$ & $h$ & $R$ & $b$ & $\rho$  &$f_1\,\, (E_1)$
& $f_2\,\, (E_2)$ & $f_3\,\, (E_3)$ & $f_4\,\, (E_4)$ & $f_5\,\, (E_5)$ \\
\tableline
0.2  & 0.0  &  0.7 & 2.0  & 0.0 & 8.1 & 13 ~(4.0) & 5.3 ~(3.2) & 3.3
{}~(2.7) & 2.4 ~(2.4) & 1.9 ~(2.2)\\
1.0  & 0.0  &  0.7 & 0.9  & 0.0  & 8.4 & 1.6 ~(3.4) & 1.1 ~(3.2) &
0.9 ~(3.0) &  0.8 ~(2.9) &  0.7 ~(2.8)\\
\tableline
 0.15  & 0.2   &  0.7 & 2.4   & 13  & 5.0  & 4.0 ~(3.1) & 5.4 ~(2.3)
& 5.3
{}~(1.8) & 4.9 ~(1.5) & 4.0 ~(1.3)\\
 0.2   & 0.1   &  0.7 & 2.0   & 4.5  & 7.6 & 7.4 ~(3.8) & 7.2 ~(3.0)
& 4.9
{}~(2.5) & 3.2 ~(2.2) & 2.4 ~(2.0)\\
 "   & 0.2   &  0.7 & 2.0  & 12  & 5.4 & 3.1 ~(3.4) & 4.1 ~(2.6) &
4.3 ~(2.1) &
4.1 ~(1.8) & 3.6 ~(1.6)\\
 "   & "   &  0.5 & 2.8 & 17 & 4.9 & 1.9 ~(3.5) & 2.5 ~(2.7) & 2.8
{}~(2.3) & 2.9
{}~(2.0) & 3.0 ~(1.7)\\
 "   & "   &  0.9 & 1.6  & 9.3 & 6.0 & 4.4 ~(3.2) & 5.3 ~(2.5) & 5.1
{}~(2.0) &
4.5 ~(1.7) & 3.6 ~(1.5)\\
 "   & 0.4   &  0.7 & 2.0   & 40  & 2.6 &  0.8 ~(2.5) & 1.6 ~(1.8) &
2.1 ~(1.4)
& 2.4 ~(1.1) & 2.6 ~( 0.9)\\
 0.25  & 0.2  &  0.7 & 1.8 & 8.5 & 5.5 & 2.0 ~(3.5) & 2.9 ~(2.8) &
3.3 ~(2.4) &
3.4 ~(2.1) & 3.1 ~(1.8)\\
 0.4  & 0.2  &  0.7 & 1.5 & 2.2 & 7.7 & 1.1 ~(4.0) & 1.5 ~(3.4) & 1.8
{}~(3.0) &
1.9 ~(2.8) & 2.1 ~(2.5)\\
\end{tabular}
\label{tbl3}
\end{table}

\begin{table}
\caption{Values of $j_n$, $E_n$  and $\rho_n$ given by Eqs.(\ref{s5e14},
\ref{s5e16}) for the case $\epsilon=0.2$, $\bar{j}=0.2$, $h=0.7$}
\begin{tabular}{cccc}
 $n$ &  $j_n$ & $\rho_n\, (10^{-26}$ g cm$^{-3})$ & $E_n\, 
(\frac{1}{2}(300$ km s$^{-1})^2)$  \\
\tableline
 1  & 0.05  &  1.7 & 4.9  \\
 2  & 0.11  &  2.5 & 3.8  \\
 3  & 0.17  &  2.7 & 3.3  \\
 4  & 0.22  &  2.5 & 2.9 \\
 5  & 0.26  &  2.2 & 2.6  \\
 6  & 0.30  &  1.9 & 2.4  \\
 7  & 0.34  &  1.6 & 2.2  \\
 8  & 0.38  &  1.4 & 2.0  \\
\end{tabular}
\label{V}
\end{table}

%%%%%%%%%%%%%%%%%%%%%%

\def\baselinestretch{.7}

\end{document}